\documentclass[lettersize,magazine]{IEEEtran}
\usepackage{amsmath,amsfonts,amssymb}
\usepackage{algorithmic}
\usepackage{algorithm}
\usepackage{array}
\usepackage[caption=false,font=normalsize,labelfont=sf,textfont=sf]{subfig}
\usepackage{textcomp}
\usepackage{stfloats}
\usepackage{url}
\usepackage{verbatim}
\usepackage{graphicx}
\usepackage{cite}
\usepackage{color}
\PassOptionsToPackage{table,dvipsnames}{xcolor}
\usepackage[table,dvipsnames]{xcolor}
\usepackage{xcolor}
\usepackage{booktabs}
\usepackage{arydshln}
\usepackage{longtable}
\usepackage{caption}
\usepackage{hyperref}
\usepackage{multirow}

\usepackage[thmmarks, thref, amsmath]{ntheorem}
\theoremstyle{plain}
\theoremprework{\bigskip\hrule\vspace{-1.5mm}\leavevmode\nobreak}
\theorempostwork{\vspace*{-1mm}\hrule\bigskip\leavevmode}
\theoremheaderfont{\scshape}
\theorembodyfont{\itshape}
\theoremseparator{. }

\newtheorem{remark}{Remark}
\newtheorem{example}{Example}

\begin{document}
\title{Foundations for Energy-Aware Zero-Energy Devices: From Energy Sensing to\\ Adaptive Protocols}

\author{Onel L. A. López,~\IEEEmembership{Senior Member,~IEEE}, Mateen Ashraf, Samer Nasser,~\IEEEmembership{Student Member,~IEEE}, Gabriel M. de Jesus,~\IEEEmembership{Graduate Student Member,~IEEE}, Ritesh Kumar Singh,~\IEEEmembership{Member,~IEEE}, Miltiadis C. Filippou,~\IEEEmembership{Senior Member,~IEEE}, Jeroen Famaey,~\IEEEmembership{Senior Member,~IEEE}
\thanks{Onel López, Mateen Ashraf, and  Gabriel M. de Jesus are with the Centre for Wireless Communications (CWC), University of Oulu, Finland. \{onel.alcarazlopez,mateen.ashraf,gabriel.martinsdejesus)@oulu.fi.}
\thanks{Samer Nasser, Ritesh Kumar Singh, and  Jeroen Famaey are with the IDLab, University of Antwerp – IMEC, Belgium. \{samer.nasser, ritesh.singh, jeroen.famaey)@imec.be.}
\thanks{Miltiadis C. Filippou is with WINGS ICT Solutions S.A., Athens, Greece. mfilippou@wings-ict-solutions.eu}
\thanks{This work has been partially supported by the Research Council of Finland (Grants 362782 (ECO-LITE) and 369116 (6G Flagship)) and the European Commission through the Horizon Europe/JU SNS project AMBIENT-6G (Grant 101192113).}
}

\maketitle

\begin{abstract}
Zero-energy devices (ZEDs) are key enablers of sustainable Internet of Things networks by operating solely on harvested ambient energy.
Their limited and dynamic energy budget calls for protocols that are energy-aware and intelligently adaptive. However, designing effective energy-aware protocols for ZEDs requires theoretical models that realistically reflect device constraints. Indeed, existing approaches often oversimplify key aspects such as energy information (EI) acquisition, task-level variability, and energy storage dynamics, limiting their practical relevance and transferability. 
This article addresses this gap by offering a structured overview of the key modeling components, trade-offs, and limitations involved in energy-aware ZED protocol design. For this, we dissect EI acquisition methods and costs, characterize core operational tasks, analyze energy usage models and storage constraints, and review representative protocol strategies. 
%Moreover, we offer design insights and guidelines on how the ZED tasks can actively leverage EI to improve performance, in several cases aided by in-house examples, indicate open research directions aimed at guiding more grounded and adaptable protocol development for ZED platforms.
Moreover, we offer design insights and guidelines on how ZED operation protocols can leverage EI, often illustrated through selected in-house examples. Finally, we outline key research directions to inspire more efficient and scalable protocol solutions for future ZEDs.
\end{abstract}


\section{Introduction}\label{intro}
Sustainability is becoming a central pillar in the design of next-generation wireless communication systems, reflecting broader societal efforts for balanced economic development, social equity and well-being, and environmental protection \cite{Lopez.2023,Rahmani.2023}.
Among these systems, the Internet of Things (IoT) stands out due to its sheer scale and transformative potential \cite{Mahmood.2020}, with approximately 39 billion connected devices projected by the end of 2029, which may represent a 147\% increase from 2023, according to \href{https://www.ericsson.com/en/reports-and-papers/mobility-report/dataforecasts/iot-connections-outlook}{Ericsson}. 
Note that the IoT plays a dual role: not only must IoT promote/enable sustainable practices across a multitude of sectors/domains, such as energy management, precision agriculture, smart mobility, and environmental monitoring, but its infrastructure must also align with sustainability goals \cite{Banotra.2023,Lopez.2023}. Indeed, as IoT networks grow in scale, issues like 
increased maintenance operations and manufacturing/disposal of IoT devices become a pressing concern, motivating the development and increasing adoption of energy harvesting (EH) technologies and techniques \cite{Rahmani.2023,LopezAlves.2021,Calautit.2021}.

EH refers to capturing and converting ambient energy into usable electrical energy to power electronic systems within a hosting device \cite{LopezAlves.2021}. For IoT devices, EH-based charging is a compelling alternative to traditional battery-based (or even wired) charging, as it may i) enable devices to operate autonomously over extended periods with minimal maintenance, increasing their durability thanks to the contact-free feature; ii) enhance deployment flexibility, e.g., facilitating IoT applications in hazardous environments, building structures, or the human body; and iii) promote environmental sustainability by reducing the network-wide emissions footprint and hazardous/pollutant electronic waste processing \cite{LopezAlves.2021,Alves.2021}. These IoT devices are commonly referred to as ambiently-powered, energy-neutral, or zero-energy devices (ZEDs) \cite{hexaX2.2023,Lopez.2025}, specially when EH is the only charging source. Herein, we adopt the latter, most popular notation.

ZEDs may exploit varied EH sources, each with distinct characteristics, especially in terms of availability, power density, and required transducers \cite{Lopez.2025,Unlu.2018,Akan.2018,Smart.2016,Sanislav.2021,Leemput.2023}. 
The most common are i) light, which provides high power density but depends on illumination conditions; ii) thermal/temperature gradients, often suitable for industrial settings; iii) kinetic energy from motion or vibration, captured using piezoelectric or electromagnetic mechanisms and often applied in wearable or mobile systems; and iv) radio frequency (RF) EH, which is particularly attractive for ultra-low-power and miniaturized devices. Notably, piezoelectric (for motion/vibration EH) and photovoltaic (for light EH) transducers are the most versatile and commercially mature technologies nowadays \cite{Smart.2016,Sanislav.2021,Leemput.2023}.

Despite its promise, integrating EH into IoT devices introduces several inherent challenges that complicate system design and operation.
On the one hand, incorporating EH hardware (HW) may somewhat increase design complexity and cost, and affect device size, form factor, and deployment options, depending on the specific EH source/transducers and envisioned applications \cite{Lopez.2023,Rahmani.2023}. 
On the other hand, the availability and strength of ambient energy sources are often limited and variable \cite{hexaX2.2023,Alves.2021,Lopez.2025,Unlu.2018}, which, together with low EH efficiencies and energy-storage limitations \cite{Lopez.2023,Calautit.2021,Zeadally.2020}, especially in miniaturized harvesters, make it difficult to guarantee consistent device performance. For instance, regarding RF-EH, which can be integrated in very small form factors, the effectiveness of harvesting is strongly influenced by the number and parameters of the RF transmitters, their distance to the ZED(s), and the operation frequency. Specifically, fewer transmitters, greater distances, and higher frequencies lead to less radiated power, higher path loss, and smaller receive apertures, respectively, thus lower efficiencies. 
All in all, and due to all this, ZEDs often operate intermittently \cite{Ma.2020,Sumanth.2021,Islam.2020,Min.2022,Sandhu.2021,Karimi.2023}.

\begin{figure}[t!]
    \centering    \includegraphics[width=\linewidth]{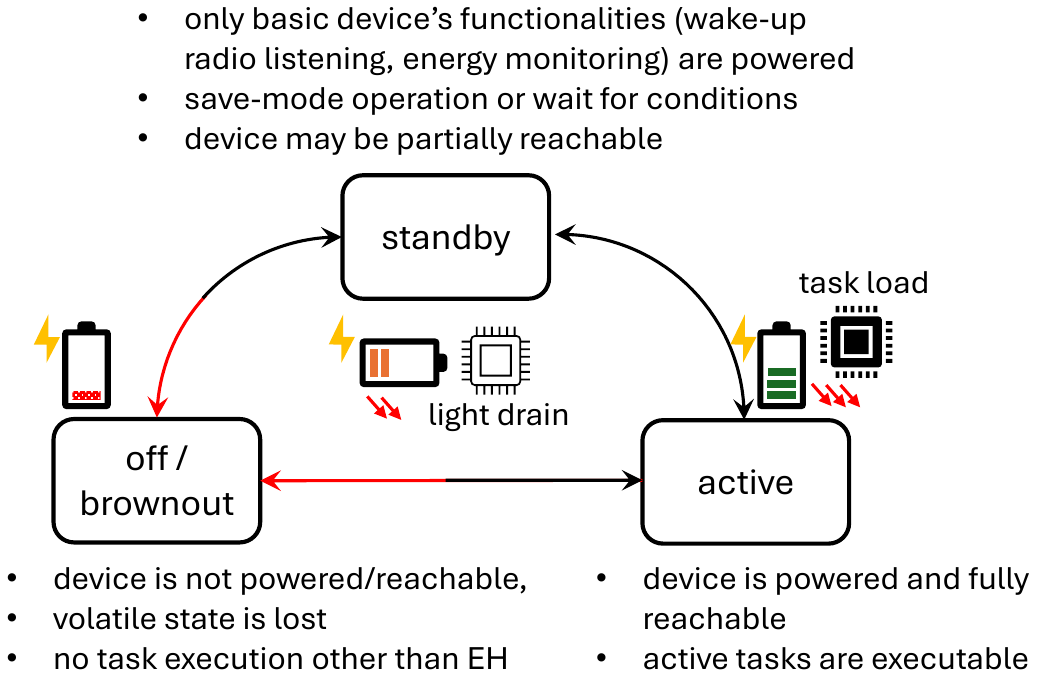}
    \caption{High-level operation phases of ZEDs and their key features.}
    \label{fig:states}
\end{figure}

The operation of ZEDs can be abstracted into three high-level phases or states, namely off/brownout, standby, and active \cite{Lee.2013,Sandhu.2021,Karimi.2023}, as illustrated in Fig.~\ref{fig:states}. In the off/brownout phase, the device lacks sufficient energy to remain powered; hence, all execution halts, and only EH continues passively until the storage recharges sufficiently. Once the energy reaches a usable threshold, the device enters the standby or active phase. In standby, the device remains powered but idle or minimally active, potentially monitoring energy conditions, checking timers or interrupts, and preparing for execution. Herein, energy-intensive components, such as the microcontroller unit (MCU), sensors, and radio units, are either powered down or transitioned into sleep mode. In the case of embedded devices, only the timekeeping ability, e.g., MCU's clock unit or external programmable real-time clock, may remain active \cite{Karimi.2023,Islam.2020}.\footnote{There may be different standby modes, including power-down and power-save \cite{Shnayder.2004}.} The device transitions to active (from either off/brownout or standby phase) and performs its tasks when sufficient energy is available and conditions are met.

For ZEDs to function dependably within tight energy budgets, their design must follow minimalistic system principles, aiming to reduce memory usage, computational complexity, and peripheral activity at both the HW and software (SW) levels \cite{Rahmani.2023,Calautit.2021,Lopez.2025}. 
Whenever possible/applicable, they must also sustain task/operation progress despite frequent power interruptions \cite{Hester.2017,Maeng.2019,Maeng.2020,Ma.2020,Sumanth.2021,Min.2022}. This is achieved, e.g., by preserving the system state in non-volatile memory before power loss and resuming execution upon energy recovery. 
Complementing these, adaptive operation mechanisms are essential for aligning system behavior with the fluctuating energy conditions, achieving long-term operation without manual intervention. Such mechanisms must incorporate energy awareness into core decision-making processes, enabling tasks to be deferred, adjusted, or triggered based on energy information (EI),\footnote{EI may refer to different quantities and be acquired correspondingly in several ways, as discussed later in Section~\ref{sec:acqusition}.}
thereby ensuring both responsiveness and resilience in energy-constrained environments \cite{Sabovic.2020,hexaX2.2023,Karimi.2022,Delgado.2022,Lopez.2023,Adnan.2023,Valente.2023,Giannini.2025,Gerhorst.2020}.
Herein, we precisely focus on energy-aware operation as a foundational paradigm for enabling dependable ZED systems and/or extending their application horizons.
\subsection{State of the Art}
The research literature on energy-aware/adaptive operation protocols for ZEDs is immense.\footnote{Indeed, a Google Scholar search indicated 
more than 8000 papers from 2018 onward with the keywords ``energy harvesting'' and ``IoT'' and at least one out of the keywords ``energy-aware'' and ``energy-adaptive''. A deeper look into the search made us realize that about $35\%$ of these works really deal with energy-aware/adaptive ZED protocols, which is still a huge number, and that most of the literature is from recent years, with only 11\% of the works authored before 2018.} 
Broadly, we can classify these efforts into three categories, which, together with the corresponding state-of-the-art, are discussed below and captured in Table~\ref{tab:SoTA}.
\subsubsection{Purely Theoretical Studies}
Theoretical studies have primarily focused on high-level modeling and protocol development. They typically study the fundamental trade-offs between energy availability, task execution, and application performance, always without building physical prototypes.
These studies pose the ZEDs' operation under energy constraints as mathematical optimization problems, often proposing heuristics \cite{Valente.2023,Sangrez.2021,Sarang.2023,Giannini.2025}, robust (sub)optimal policies \cite{Lopez.2022,Delgado.2022}, or performance bounds \cite{Lopez.2022}. 
The optimization focus is varied, ranging from maximizing medium access control (MAC) success probabilities \cite{Sangrez.2021,Sarang.2023}, the number of executed tasks \cite{Delgado.2022}, model learning convergence \cite{Valente.2023}, and harvested ambient RF energy, to minimizing the network's age of information (AoI) \cite{Giannini.2025}. 

 Though many of these works are foundational, they are increasingly struggling to capture the true operational constraints and characteristics of ultra-constrained ZEDs, such as complex EH and energy storage dynamics \cite{Valente.2023,Giannini.2025,Sangrez.2021}, limited memory/processing/communication capabilities \cite{Valente.2023,Sangrez.2021,Lopez.2022,Sarang.2023,Giannini.2025}, non-deterministic load behavior \cite{Valente.2023,Sangrez.2021,Sarang.2023,Giannini.2025}, and frequent energy interruptions \cite{Valente.2023,Sangrez.2021,Lopez.2022,Sarang.2023,Giannini.2025}. Indeed, theoretical models tend to abstract away these details to achieve tractable formulations. However, as a result, they often overestimate performance, or worse, incorrectly claim protocol feasibility. 
 Unlike traditional systems, where a model's inaccuracy degree may still yield valuable insights, the tight margins of ZEDs make even small modeling inaccuracies result in unworkable protocols. Consequently, theoretical studies risk becoming disconnected from the realities they aim to guide unless they are recalibrated with accurate operation models.
 \begin{table*}[t!]
    \centering
    \caption{Some representative works on energy-aware ZED operation in the period 2018-2025}
    \label{tab:SoTA}
    \begin{tabular}{p{0.15cm}|p{0.3cm}p{2.6cm}p{2.6cm}p{2.7cm}p{3.6cm}p{3.9cm}}
    \toprule
        & \textbf{ref.} & \textbf{setup}  & \textbf{energy-awareness for} & \textbf{tasks}  & \textbf{key insights} & \textbf{key limitations} \\
        \midrule
        \multirow{10}{*}{\rotatebox[origin=c]{90}{purely theorethical}}
         & \cite{Delgado.2022} & battery-less ZED
         & task and sleep scheduling  & sensing, processing, \& reporting  & how much EH look-ahead time for optimal performance
         & known task energy consumption and future EH   \\ \cdashline{2-7}
         & \cite{Valente.2023} & federated learning network &  probabilistic sleeping/engagement & model learning \& reporting  & average EI sharing leads to near-optimal performance & simplistic battery dynamics and tasks (and energy consumption) \\ \cdashline{2-7}
         & \cite{Giannini.2025} & monitoring network & probabilistic sleeping/engagement  & reporting  & joint energy \& AoI awareness helps minimize network AoI & simplistic battery and task dynamics; full data buffer  \\ \cdashline{2-7}
         & \cite{Sangrez.2021} & personal area network with RF-EH from LTE & multiple access communication and control & transmission & IEEE 802.15.4 may benefit from energy-awareness & simplistic battery and task dynamics; full data buffer  \\ \cdashline{2-7}
         & \cite{Sarang.2023} & receiver-initiated communication network & duty-cycling & communication (different transceiver states) & (learned) future EI helps stabilize network dynamics  & abstracted tasks; ideal (overhead-free) protocols/learning \\ \cdashline{2-7}
         & \cite{Lopez.2022} &  multi-antenna dynamic RF-EH combiner & RF receive beamsteering & no tasks (only EH) & EI is useful in (relatively) high RF-EH rate scenarios & narrowband RF-EH; simplistic energy storage dynamics \\
         \midrule
       \parbox[t]{2mm}{\multirow{9}{*}{\rotatebox[origin=c]{90}{prototype-based}}}         
         & \cite{Maeng.2019}  & ZED with volatile \& non-volatile memory & dynamic peripheral workload scaling  & peripheral operations & scaling peripheral tasks is advisable when plausible &    no concurrent computation and peripheral access    
         \\ \cdashline{2-7}
         & \cite{Adnan.2023}  & battery-less ZED with cloud support & local vs cloud intelligence decision & processing (machine learning) & local inference is preferred in controllable EH setups & periodic local data acquisition; specific user application \\ \cdashline{2-7}
         & \cite{Colin.2018} & reconfigurable energy storage system & storage/capacity reconfiguration & generic (atomic and/or reactive) & reconfigurable storage supports varied tasks efficiently & task profiling is fixed and specified beforehand\\ \cdashline{2-7}
         & \cite{Mosch.2020} & reconfigurable vibration transducer & matching/tuning to vibration frequency & no tasks (only EH) & accurate dominant frequency tuning demands tens of $\mu$J & only a transducer; tests in ideal conditions  \\ \cdashline{2-7}
         & \cite{Rajappa.2023} & separate PMU$^\dagger$ and interfaces to a ZED & task scheduling & sensing \& (abstract) transmission chain & current non-energy-aware can become energy-aware ZEDs & higher complexity that may not compensate the gains \\ \cdashline{2-7}
         & \cite{Yang.2019} & battery-less ZED & scheduling task execution rates & generic & heavy-duty-cycled storage voltage monitoring may suffice &  independently-controlled tasks without priorities \\   \cdashline{2-7}
         & \cite{Yildirim.2018} & kernel for battery-less ZEDs & reactive task scheduling & event-driven sensing, processing, \& reporting  &  advanced systems call for inter-task thread communication, event handling, \& timekeeping & pre-defined/programmed task work-flow; no guarantees for peripheral state consistency \\
         \midrule
         \parbox[t]{2mm}{\multirow{4}{*}{\rotatebox[origin=c]{90}{hybrid}}} 
         & \cite{Islam.2020} & intermittent MCU-based ZED & on/off-line scheduling tasks within deadlines  & generic & deadline violations may decrease greatly, but not avoided & periodic tasks arrival; fixed storage charging rate   \\ \cdashline{2-7}
         & \cite{Karimi.2023} & real-time operating system for battery-less ZEDs & task scheduling and checkpointing & computational and peripheral tasks & mixed-preemption scheduling may facilitate the execution of
multi-task applications & fixed EH rate and tasks' charging demands\\   \cdashline{2-7} 
         & \cite{Maeng.2020} & ZED with volatile \& non-volatile memory  & task scheduling & generic (periodic and reactive)  & distinguishing tasks from time-sensitive to insensitive is crucial &  pre-defined/configured tasks; single application  \\   \cdashline{2-7}      
         & \cite{Sabovic.2020} & battery-less LoRa ZED & (off-line) duty-cycling 
         &  sensing \& (abstract) transmission chain  &  capacitor size sets the needed EH rate \& full cycle time & non-adaptive; limited to constant-current EH sources  \\ \cdashline{2-7}
         & \cite{Geissdoerfer.2022}& battery-less ZED to ZED link & maximizing rendezvous chances & wake-up and link establishment & charging times $\sim$follow well-known probability distributions & independently-distributed ZEDs' charging times  
         \\ \cdashline{2-7}
          & \cite{Jewsakul.2025} & LoRa network & probabilistic sleeping/engagement & transmission & $\sim\!+\!100\%$ lifetime due to current/predicted EI leverage & not a full testbed: some components used/tested offline  \\ \cdashline{2-7}
         & \cite{Ibrahim.2021} &  reconfigurable vibration transducer & matching/tuning to vibration frequency & no tasks (only EH) & power required to overcome motor friction only & only a transducer; large form factor \\            
         \bottomrule
    \end{tabular} 
    \footnotesize
    \begin{flushleft}
        $^\dagger$ PMU: power management unit
    \end{flushleft}
\end{table*}

\subsubsection{Prototype-based Studies}
These studies emphasize real-world implementations, building HW/SW component prototypes \cite{Colin.2018,Mosch.2020,Rajappa.2023,Adnan.2023}, kernels \cite{Maeng.2019,Yang.2019,Yildirim.2018}, or even full ZED prototypes or platforms \cite{Yang.2019} to validate energy-aware protocols under actual conditions. They typically revolve around specific applications and/or energy sources. Common themes include the implementation of intermittent computing systems \cite{Maeng.2019,Yildirim.2018} and adaptive operation patterns tailored to energy available/consumption profiles \cite{Colin.2018,Maeng.2019,Yildirim.2018}, conditions \cite{Yang.2019,Mosch.2020,Rajappa.2023,Adnan.2023,Yildirim.2018}, or predictions \cite{Rajappa.2023,Adnan.2023}. 

However, these systems are often narrow in focus, tightly bound to specific HW configurations \cite{Colin.2018,Mosch.2020,Rajappa.2023,Yildirim.2018}, energy environments \cite{Mosch.2020,Adnan.2023}, and application requirements \cite{Maeng.2019,Yildirim.2018}. As such, the insights they offer are not easily generalizable to other platforms, conditions, or use cases, even with slightly different parameter setups. Just as an example, the adaptive control strategies in \cite{Colin.2018} use fixed design parameters, such as capacitor size, sampling intervals, and energy thresholds, without exploring how these choices impact performance under varying conditions or system configurations. Furthermore, many prototype studies lack formal modeling, which makes it difficult to predict how observed behaviors scale or change under different conditions. Although platform-agnostic frameworks like AsTAR \cite{Yang.2019} aim to mitigate these constraints, they
may hinder precise modeling and performance assessment, as they abstract away hardware-specific timing, energy behaviors, and execution semantics, which are crucial for proper optimization. 

In general, the usual focus is on making the system ``work'' rather than systematically analyzing why a particular configuration is successful and/or identifying key trade-offs to optimize performance further, leaving gaps in formal understanding and design principles.
This limits their value in shaping general design principles or informing theoretical frameworks.

\subsubsection{Hybrid Approaches}
Hybrid studies combine theory frameworks with prototyping realism, often developing mathematical models or algorithms and validating them experimentally. These works have introduced energy-aware/adaptive kernels \cite{Maeng.2020,Islam.2020,Karimi.2023}, voltage operation thresholds \cite{Sabovic.2020,Karimi.2023}, reactive schedulers \cite{Geissdoerfer.2022,Islam.2020,Jewsakul.2025}, and adaptive runtime strategies \cite{Islam.2020,Ibrahim.2021,Jewsakul.2025,Karimi.2023} that are both theory-justified and demonstrated on ZED platforms. Examples include systems like Celebi \cite{Islam.2020}, an offline/online scheduler for computation tasks with deadline constraints; CARTOS \cite{Karimi.2023}, a charging-aware real-time IoT operating system supporting task chains processing; CatNap \cite{Maeng.2020}, an event-driven scheduler relying on predefined binary execution priorities; Bonito \cite{Geissdoerfer.2022}, a connection protocol for reliable bi-directional communication between intermittent ZEDs; RACEME \cite{Jewsakul.2025}, an EH-management framework leveraging the spatio-temporal correlation between ZEDs and their embedded intelligence to optimize the harvested-energy utilization.

Although unifying theory and practice, these approaches are not without limitations. In many cases, the theory remains at a high level while the experimental validation is narrow in scope, often limited to specific energy traces or controlled and fixed settings. Indeed, the theoretical model components still often rely on critical assumptions given ultra-constrained ZED contexts, such as ideal/stationary energy prediction and/or energy availability models that abstract away fine-grained fluctuations and conversion losses \cite{Islam.2020,Geissdoerfer.2022,Jewsakul.2025,Karimi.2023}, under-modeling of system/device-level overheads \cite{Jewsakul.2025,Maeng.2020,Karimi.2023}, isolated HW/SW component designs    \cite{Ibrahim.2021,Jewsakul.2025,Karimi.2023}, or simplified task models \cite{Islam.2020,Sabovic.2020,Maeng.2020,Geissdoerfer.2022,Karimi.2023}, that don't fully reflect the constraints of real-world systems. As a result, even hybrid studies can fall short of providing robust, generalizable frameworks for energy-aware ZED operation.
\subsection{Motivation and Contributions}
One pressing need in energy-aware ZED protocol design is to improve the realism of theoretical models with respect to actual ZED behavior. As stated earlier, accurate theoretical models are crucial to enable systematic exploration of design trade-offs, scalability analysis, and performance bounds across diverse conditions, something that prototype-based studies alone cannot provide due to their narrow scope and limited generalizability. 
Aspects like the energy/time cost and accuracy of EI acquisition, the variability in task-level energy and timing characteristics, and the nonlinear dynamics of energy storage components must be properly modeled for obtaining meaningful insights and designing workable ZED operation protocols. To guide future research in this direction, this work overviews models, trade-offs, and key challenges involved in making energy-aware ZED protocol design more realistic. Our specific contributions are five-fold:
\begin{itemize}
    \item We provide a structured breakdown of EI acquisition in ZEDs, covering the various types of EI relevant to energy-aware protocols and identifying where and how this information can be obtained. This includes a categorization of measurement points and acquisition methods, including forecasting, and their associated overheads.
    Notably, such overheads are rarely accounted for in the purely-theoretical literature, with exceptions like \cite{Lopez.2022}, while they are relatively more common in works including prototyping, such as \cite{Islam.2020,Maeng.2019,Maeng.2020,Sabovic.2020,Adnan.2023,Colin.2018,Rajappa.2023,Yang.2019,Geissdoerfer.2022,Jewsakul.2025}, but in all the cases, the treatment is system-specific and also rather superficial and narrow. These aspects are examined in detail in Section~\ref{sec:acqusition}.   
    \item We characterize the core operation tasks in ZEDs, namely sensing, computation, communication, and actuation. Unlike existing overviews/surveys such as \cite{Zeadally.2020,Calautit.2021,Sumanth.2021,Rahmani.2023}, which discuss specific tasks alone, and \cite{Ma.2020,Sandhu.2021,Min.2022,Banotra.2023}, which dilute the discussions within broad overviews of EH-IoT architectures, applications, and system-level behaviors, our treatment isolates and analyzes the practical constraints of these core tasks. Specifically, our analysis focuses on the tasks' energy cost profiles, identifying typical consumption ranges and influencing factors; execution granularity, highlighting how tasks are structured into energy-compatible segments; and response timeliness requirements, which affect the feasibility of energy-aware adaptation under intermittent power. This task-level view, detailed in Section~\ref{sec:tasks}, offers a sharper lens to modeling and/or considering actual ZED capabilities.    
    \item We overview energy evolution and usage models, emphasizing the different EH source and load behaviors and how energy constraints manifest and can be enforced at different granularities. Moreover, we examine the limitations imposed by energy storage components and how these physical constraints shape the feasible space for energy-aware control. These aspects are discussed in Section~\ref{sec:usage}, while very rarely and shallowly in the literature, which focuses more on implementation-specific discussions \cite{Delgado.2022,Sarang.2023,Valente.2023,Giannini.2025,Lopez.2022,Adnan.2023,Mosch.2020,Rajappa.2023,Sabovic.2020,Jewsakul.2025,Ibrahim.2021} or broad system-level energy dynamics characterization and scheduling \cite{Ma.2020,Sandhu.2021,Min.2022,Valente.2023,Giannini.2025,Sangrez.2021,Sarang.2023,Maeng.2019,Adnan.2023,Colin.2018,Yang.2019,Islam.2020,Maeng.2020,Geissdoerfer.2022} and energy storage and EH technologies overviewing \cite{Calautit.2021,Sumanth.2021,Banotra.2023}. Our work here highlights the need for capturing the real dynamics of energy availability and usage in ZEDs and provides specific pointers for further research.
    \item We offer design insights and guidelines on how the ZED tasks can actively leverage EI to improve performance. This is grounded in a structured review of representative energy-aware protocols from the literature \cite{Islam.2020,Maeng.2019,Maeng.2020,Sabovic.2020,Delgado.2022,Adnan.2023,Valente.2023,Giannini.2025,Sangrez.2021,Sarang.2023,Lopez.2022,Colin.2018,Mosch.2020,Rajappa.2023,Yang.2019,Geissdoerfer.2022,Jewsakul.2025,Ibrahim.2021,Karimi.2023,Yildirim.2018}, including the few considering EI acquisition overheads, and several in-house examples. These discussions are presented in Section~\ref{sec:protocols}.
    \item We identify key challenges and research directions  throughout the paper, culminating in a consolidated summary of open issues and actionable insights in Section~\ref{sec:conclusion}.   
\end{itemize}
Through the discussions and insights provided throughout, we aim to inspire efforts toward more realistic, efficient, and adaptable energy-aware protocol designs for ZED systems.
\section{EI Acquisition \label{sec:acqusition}}
\begin{figure}[t!]
    \centering
    \includegraphics[width=1\linewidth]{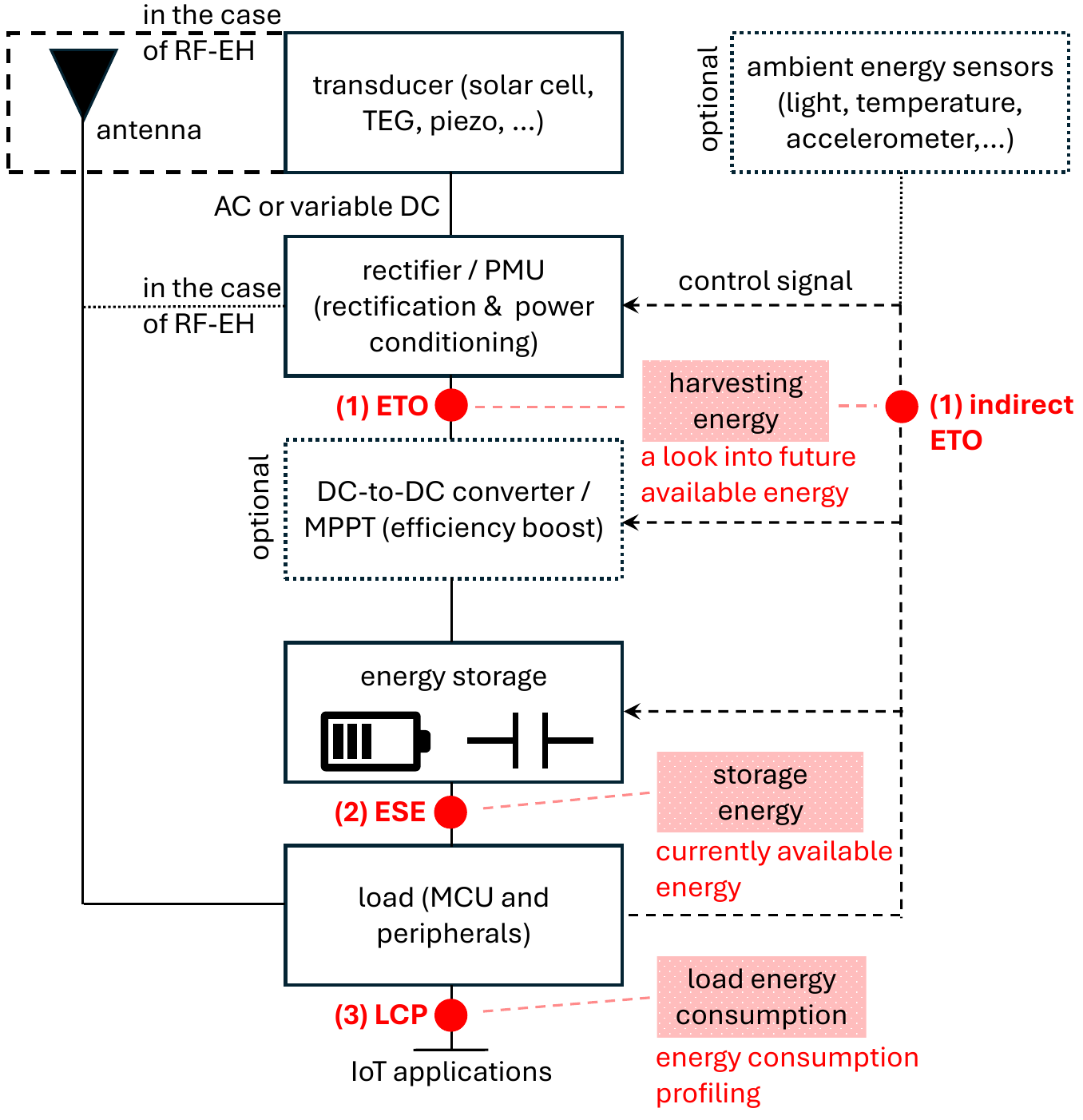}
    \caption{Block diagram of a generic ZED architecture. Possible EI acquisition points are indicated: energy transducer output (ETO), energy storage element (ESE), and load or device consumption points (LCP).}
    \label{fig:measurementPoints}
\end{figure}

The block diagram of a generic ZED is illustrated in Fig.~\ref{fig:measurementPoints} together with the possible EI acquisition points. The main building blocks include a transducer, rectifier, PMU, direct current (DC)-to-DC converter, maximum power point tracking (MPPT), energy storage, and load.

The \emph{transducer} converts ambient energy into electrical energy. Its output characteristics, e.g., AC/DC, voltage level, and impedance, depend heavily on the energy source \cite{Banotra.2023,Lopez.2025,Unlu.2018,Akan.2018,Smart.2016,Sanislav.2021,Leemput.2023,Ma.2020}. For instance, piezoelectric and RF transducers produce AC signals, requiring rectification, while solar cells output DC with source-dependent impedance and non-linear power profiles. Some more sophisticated ZEDs may include multiple transducers to capture energy from different ambient sources, thereby improving the statistics of the harvested energy \cite{Akan.2018,Lopez.2023}. The transducer connects to the \emph{rectifier} and \emph{PMU}, which constitute the front-end energy interface, ensuring that raw harvested energy is converted into a usable and stable supply for the rest of the system. The rectifier is used for DC conversion when the transducer outputs AC energy, while the PMU regulates the DC energy flow, typically handling cold-start conditions, protecting against over/under-voltage, and routing energy to either storage, load, or both via proper signaling \cite{Akan.2018,Lopez.2023,Ma.2020,Sandhu.2021}. In simple EH setups, this block may be just a diode and a capacitor, while it may integrate control logic, telemetry, and charge pumps in more complex systems. Next in the flow, there might be a \emph{DC-to-DC converter} and \emph{MPPT}, which optimize power transfer between the energy source, storage, and load by adapting voltage and power levels to system requirements \cite{Banotra.2023,Akan.2018,Sandhu.2021}. Specifically, the DC-to-DC converter adjusts the harvested voltage to a stable level suitable for either the load or the storage element, enabling efficient operation across a wide input voltage range and supporting voltage scaling for ultra-low-power digital subsystems. Meanwhile, the MPPT circuit dynamically controls the operating point of the transducer to extract the maximum possible power, which is critical for non-linear sources like solar panels or thermoelectric generators. MPPT may be implemented within the PMU or as a separate control loop and typically interacts with the DC-to-DC stage to regulate the input impedance seen by the transducer \cite{Ma.2020,Lopez.2023,Sandhu.2021}.  These boosting blocks or the rectifier directly connect to the \emph{energy storage}, which buffers the harvested energy. Common types of energy storage are capacitors, supercapacitors, or rechargeable batteries, each with its distinctions \cite{Banotra.2023,Calautit.2021,Zeadally.2020}. Some configurations prioritize fast energy access, while others rely on high-retention storage for sparse EH scenarios. Finally, the stored energy powers the \emph{load}, which comprises the MCU, when present, and the energy-consuming peripherals, such as sensors, actuators, radios, and memory interfaces \cite{Ma.2020,Lopez.2023,Sandhu.2021}. The load is responsible for executing functional and application-level tasks. Not all of these components are always present, but the specific composition depends on the design goals and application constraints. Greater energy availability and control may be generally achieved in higher-complexity designs, which may also incur increased form factors and cost.

Throughout the rest of the section, we delve further into where and how specifically EI can be acquired and the corresponding trade-offs. This is crucial for selecting appropriate methods that balance energy-awareness benefits against measurement overhead. 

\subsection{Measurement Points}\label{sec:MPo}
EI is obtained mainly from measurements at critical points within the IoT system, as discussed next and illustrated in Fig.~\ref{fig:measurementPoints}.
\subsubsection{Energy transducer output (ETO)}

Measuring energy at the output of the energy transducer after rectification\footnote{While measuring directly at the transducer output (before rectification) is technically possible, the generated voltage/current is typically AC or irregular. This requires more complex, energy-consuming measurement circuitry, making it generally unsuitable for low-complexity ZEDs \cite{Sandhu.2021}.} provides direct and stable insights into how effectively energy is being harvested. This measurement facilitates real-time evaluation of harvesting efficiency and supports dynamic adjustments of the EH circuit, such as MPPT \cite{Lopez.2023}. However, achieving accurate and reliable measurements at this point requires careful calibration and may be challenging over long periods due to environmental variability \cite{Sigrist.2017}. As an alternative, indirect measurements may be taken using ambient energy sensors (e.g., photodiodes, thermistors, accelerometers). Indeed, they can indicate harvesting conditions without requiring direct electrical measurement \cite{Smart.2016}, significantly reducing complexity and overhead. In any case, ETO measurements alone do not directly reflect energy availability as they don't capture intermediate power conversion losses and storage/load dynamics. 

\subsubsection{Energy storage element (ESE)}

Monitoring the voltage or state-of-charge (SoC) of the energy storage element, such as a capacitor or battery, allows inference of the usable energy currently available to the ZED \cite{Sandhu.2021}. This approach does not provide real-time data on the instantaneous EH rate but instead offers retrospective insights based on past EH and energy consumption. Indeed, prior EH and energy consumption cycles, which may or may not occur simultaneously as discussed later in Section~\ref{sec:usage}, and energy storage-related impairments define energy availability at a given time. Such information can help determine whether specific tasks can be performed reliably, especially during predictable low-harvest periods, such as nighttime for solar-powered ZEDs. EI acquisition at the ESE often involves low-overhead techniques like voltage threshold detection or infrequent measurements \cite{Colin.2018,Sabovic.2020}, simplifying implementation and minimizing additional energy consumption. Note that voltage-based estimations of stored energy levels may suffer from inaccuracies due to factors like temperature variations and component aging.
\subsubsection{Load or device consumption points (LCP)} 

Measuring energy at the load or consumption points, e.g., logging how much energy each sensor reading,  transmission, and even application task uses, allows precise tracking of energy usage patterns within the ZED \cite{Gerhorst.2020}. This facilitates detailed energy budgeting, aiding in the accurate scheduling of tasks to maintain energy-neutral operation. By understanding consumption patterns, ZEDs can effectively balance energy usage against available harvested energy. Despite these advantages, this method adds complexity and overhead because dedicated, non-intrusive\footnote{This means that the measurement itself should have minimum effects on the observed system. Interested readers may refer to \cite{Gerhorst.2020,Rottleuthner.2021} for further details on the specific options for tracking the power consumption of specific components within the ZEDs, both internally and externally.} measurement circuitry and/or continuous monitoring are required \cite{Gerhorst.2020,Rottleuthner.2021}. Additionally, it offers only indirect insights into the EH or storage status, necessitating sophisticated computation and logic to effectively correlate consumption data with available energy resources.
\subsection{Measurement Methods and Overhead}\label{sec:acquisitionM}
Energy sensing may involve measuring the voltage across and current at/through the ETO, ESE, or LCP. This often involves a shunt resistor or a current sense amplifier for current, but also many EH PMUs include built-in telemetry providing signals or data about the process.
In general, embedded reference circuits and analog front-ends involved in the EI acqusision require careful design optimizations to achieve ultra-low quiescent current consumption, i.e., in the nA or pA range \cite{Lee.2013,Teng.2022}, thus minimizing energy overhead and ensuring the sensing HW remains practically invisible within the overall ZED energy budget. This typically involves specialized analog circuit techniques, such as subthreshold transistor operation, careful transistor sizing, and selecting ultra-low-leakage CMOS fabrication processes \cite{Sigrist.2017}. By biasing transistors in the subthreshold region, where they operate below conventional threshold voltages, circuits can maintain functionality while drastically reducing leakage and static current consumption. Moreover, integrating these carefully optimized analog front-ends directly within existing power-conditioning circuits or MCU peripherals can further minimize the number of external components, thereby reducing both cost and leakage current paths.

\begin{figure*}
    \centering
    \includegraphics[width=0.95\linewidth]{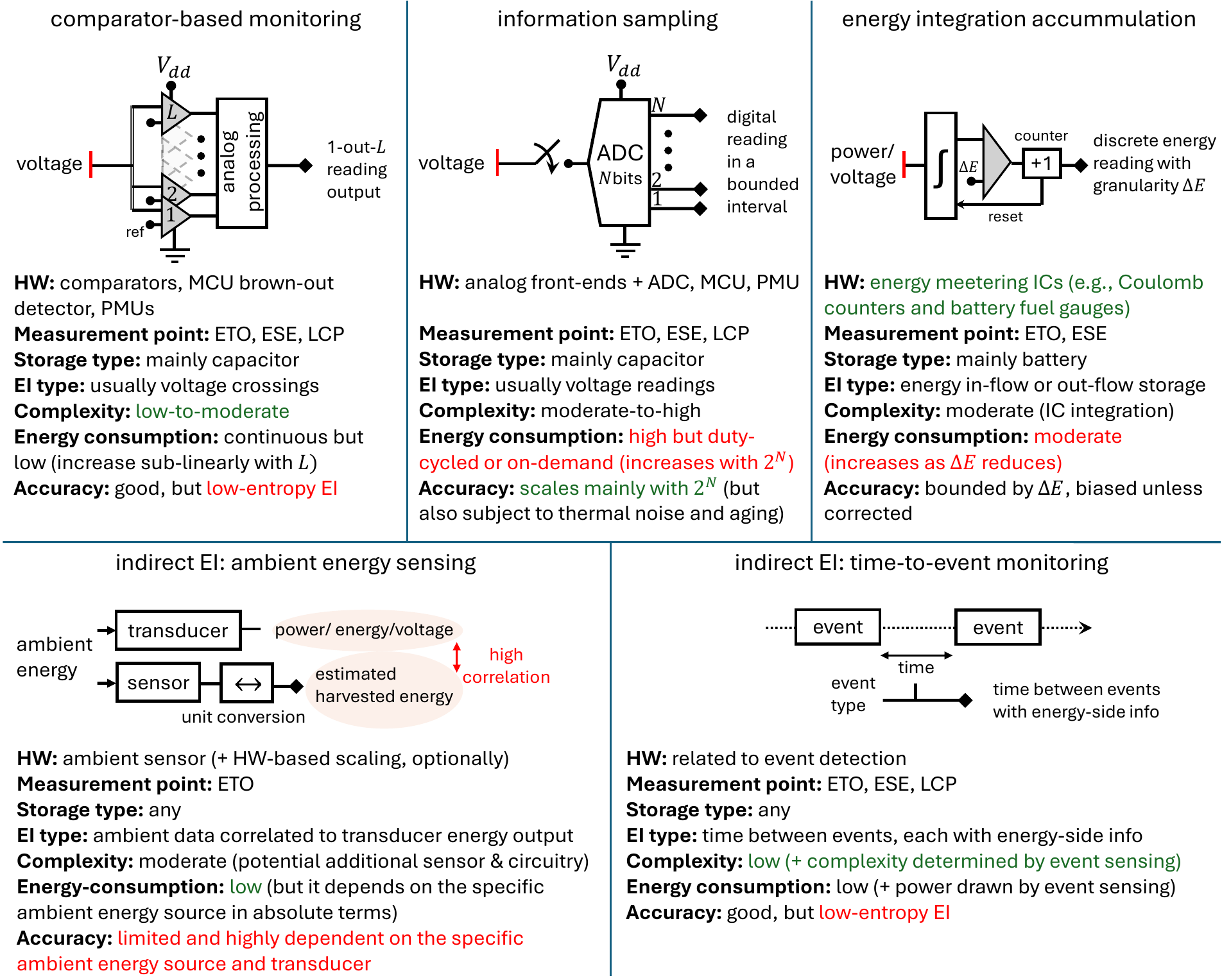}
    \caption{High-level schematics of the main EI measurement methods and their salient features. For each method, we indicate the most advantageous and disadvantageous features in green and red, respectively.}
    \label{fig:methods}
\end{figure*}

There are mainly four types of EI acquisition methods: i) comparator-based monitoring, ii) information sampling, iii) energy-integrated accumulation, and iv) indirect monitoring/sampling. Their basic principles are illustrated in Fig.~\ref{fig:methods} and discussed in detail in the following, especially focusing on corresponding HW support, energy overhead, and measurement error modeling. 
Note that in addition to these online methods, tools like source meters, oscilloscopes, and custom loggers can help profile EH performance during design and testing, thus being crucial for application/prototype development \cite{Shnayder.2004,Sigrist.2017,Eriksson.2023}. These tools are not in the final ZED but help calibrate and develop the models that might run later. For instance, by logging a week of solar panel output and light sensor readings, one can create a regression model that the ZED will use at runtime to estimate harvesting power from light \cite{Rajappa.2023}.

\subsubsection{Comparator-based Monitoring}\label{sec:compM}
Herein, there is always internal tracking using voltage threshold detection and/or low-resolution indicators, triggering a flag when a given threshold/status is reached \cite{Lee.2015,Colin.2018,Teng.2022}. For this, ultra-low-power comparators, MCU brown-out detectors, or simple on-chip threshold detectors with simple peak-holding circuits or passive rectification stages may be used. There are also PMUs with a pin that goes high when a certain voltage is reached (indicating a charged capacitor), i.e., binary energy-good signal \cite{Teng.2022,Adnan.2023,Lee.2015}, suitable for ultra-low-power implementations. In general, the comparator-based monitoring approach incurs low complexity and power overhead by providing only discrete, and often raw, EI. For instance, it can be used to indicatie when energy is available without continuous CPU running, and can be efficiently combined with PMUs, as in \cite{Lee.2015}, ensuring that the measurement overhead stays negligible and easily integrable into the EH architecture itself. 

Employing only analog comparators entails continuous passive monitoring, thus fixed but often tiny energy loss and time delay. Meanwhile, a scheduling decision may be HW/SW-triggered upon a threshold detection. This is often affordable for ultra-low-complexity ZEDs, while higher complexity ZEDs may benefit from higher-complexity/costly methods providing more granular EI.

In general, power consumption includes both dynamic and static components. The former includes switching losses due to signal fluctuations, but can often be neglected given slowly varying signals. Meanwhile the latter captures the power consumption due to leakage and quiescent current of the comparator circuit, and can be expressed as $P_c'=I_{sb} V_{dd}$ for a single comparator circuit, where $I_{sb}$ is the standby current of the comparator and $V_{dd}$ is the supply voltage. Assuming static power consumption dominates and a circuit architecture with $L$ comparators, one can write the tracking circuit power consumption as
\begin{align}
    P_c= g(L)P_c'= g(L)I_{sb} V_{dd},
\end{align}
where $g(\cdot)$, with $g(1)=1$, is a sub-linear function that depends on the specific multi-threshold comparator architecture. Indeed, the scaling with $L$ is linear in the worst case, but sub-linear scaling is usually possible by leveraging clever sharing techniques like resistor ladders and window comparators. Finally, note that as the EI here is rough, i.e., above or below a threshold, a measurement error model may lack practical usability, and we don't pursue further steps in this regard.

\subsubsection{Information Sampling}\label{sec:sampling}
Different from the previous method, the goal here is to perform explicit voltage/energy readings at specific times, e.g., as in \cite{Adnan.2023,Naderiparizi.2016,Banotra.2023}. The corresponding circuitry includes the analog front-end integrating the physical (voltage/energy) readings, e.g., a tiny shunt resistor in series with a solar panel can create a voltage proportional to current, over periods of duration $t_m$ and the ADC to provide the digital EI quantity. The activation of this circuitry is periodic or on-demand, as continuous sampling is not often possible because the energy draw is much higher now. Hence, power consumption is ``off'' most of the time, then spikes up during measurement. 

Note that many MCUs have built-in ADCs that can be used for EI sampling, while PMUs might have a connection interface to report voltages. In general, ultra-low-power communication interfaces and signaling protocols are needed, including low-speed  inter-integrated circuit ($\text{I}^2\text{C}$), serial peripheral interface (SPI), or even simple general-purpose input/output (GPIO)-based signaling between measurement integrated circuits (ICs), PMUs, and MCUs \cite{Lee.2013,Zidar.2023}. Proper design choices are crucial here. For instance, SPI provides higher speed and efficiency for high-throughput tasks but at the cost of greater wiring and energy per transaction, while $\text{I}^2\text{C}$ uses a simpler two-wire design and is usually more suitable for low-duty-cycle applications due to lower idle power and reduced complexity \cite{Zidar.2023}. Some PMUs also perform MPPT; reading their internal status can indirectly give the harvester’s operating point, from which harvested power might be inferred \cite{Adnan.2023}. Nevertheless, not all PMUs expose detailed info, and adding digital communication with a PMU can increase complexity.

The energy consumption here can be written as 
\begin{align}
E_{c}=P_{an}t_{m} + C_s V_{dd}^2,
\end{align}
where $P_{an}$ is the power consumed by the analog front-end, such as op-amps, buffers, and sample-and-hold, usually in the range of $100$ nW to $10\ \mu$W, while $C_s$ is the effective switching capacitance, and $V_{dd}$ is the supply voltage. Note that $C_s$ depends on architecture and resolution, scaling with the number of ADC bits $N$ \cite{Saberi.2011}. For example, charge-redistribution successive approximation register ADCs, common in low-power designs, use capacitor arrays, making $C_s\propto 2^N$ \cite{Saberi.2011},\footnote{The power consumption of some specific successive approximation register ADC designs may scale slower, e.g., $\propto 2^{N/2}$ in the case of binary-weighted with attenuation capacitor or even $\propto N$ in the case of split binary-weighted implementations, at the cost of increased design complexity to achieve accurate and linear performance \cite{Saberi.2011}.} while $C_s$ in flash ADCs scales worse with $N$ because $2^N-1$ comparators operate in parallel.

Meanwhile, characterizing/modeling the measurement error in this case, where readings are explicit, can be very useful to ensure operational robustness. The measurement error can be modeled as 
\begin{align}
    \epsilon = \epsilon_{t}+\epsilon_{q}+\epsilon_{0},
\end{align}
where each addend, from left to right, represents thermal noise, quantization error, and a systematic offset. 

Assuming noise is white, stationary, and ergodic, the variance of the thermal noise error is given by $\sigma_{t}^2=A/t_{m}$, where $A$ is a front-end dependent constant. For instance, for voltage tracking, it may be computed as $A=2k T R$ in the case of resistive front-ends with an equivalent input resistance $R$, wherein $T$ is the operation temperature in Kelvin, and $k$ is Boltzmann's constant \cite{Mishonov.2022}. Also, an input-referred voltage noise density $e_n$ is often specified in the datasheets of analog front-end circuits or even op-amp, while $A=e_n^2/2$ \cite{Mishonov.2022}. The thermal noise-related error component may be modeled as a zero-mean Gaussian process as long as there are no significant nonlinearities or saturation in the front-end circuitry.

The quantization error is bounded as 
\begin{align}
    -\frac{V_{r}}{2^{N+1}}<\epsilon_{q}<\frac{V_{r}}{2^{N+1}},
\end{align}
where $V_{r}$ is the input range, indicated in voltage without loss of generality. Note that if input values are uniformly distributed relative to quantization bins, then the quantization error can be modeled as a uniform random variable in the interval, hence with variance $\sigma_q^2=V_{r}^2/(3\times 2^{2N+2})$.

Finally, $\epsilon_{0}$ is mainly caused by a mismatch or imbalance in the ADC and analog front-end. It is not random and does not average out over time. Indeed, such offset may drift slowly due to temperature or aging, and thus may be modeled at time $t$ with respect to a previous time $t_0$ as
\begin{align}
    \epsilon_{0}(t)=\epsilon_{0}(t_0)+\alpha (T(t)-T(t_0)) + \beta (t-t_0),
\end{align}
where $\alpha$ is the temperature drift coefficient and $\beta$ is the time drift rate.

Together, these three components define the measurement uncertainty in sporadic sampling systems. Design trade-offs arise depending on which component dominates: longer measurement times reduce thermal noise; higher ADC resolution reduces quantization error; and offset requires calibration or compensation. In practice, $\epsilon_{q}$ dominates over $\epsilon_t$ unless very high-resolution ADCs are employed, but these are often not affordable for low-complexity ZEDs.

In general, carefully tuned sampling intervals are crucial for balancing energy savings and measurement fidelity. Indeed, strategically timed ADC (or IC) activations  can greatly reduce the overall energy burden without significantly compromising the accuracy or utility of the data obtained \cite{Rajappa.2023}. This is especially relevant at the ETO due to the typically slow-changing nature of harvested energy conditions, meaning frequent or continuous measurement often provides diminishing returns in terms of actionable insight. Its adoption at ESE and LCP is growing, but challenges like accurately capturing rapid transient events, synchronizing measurements with consumption peaks, and/or ensuring measurement reliability despite intermittent sampling persist.

\subsubsection{Energy-Integrated Accumulation}
In this case, instead of measuring instantaneous energy or voltage, the system integrates energy/charge over time, tracking the cumulative charge passing into a capacitor or battery. Energy metering ICs like coulomb counters or battery fuel gauges  are used for this, i.e., counting ``energy packets" without needing frequent sampling \cite{Sigrist.2017,Naderiparizi.2016,Rottleuthner.2021}. These HW meters offload the ADC processing from the MCU and are designed to be ultra-low-power. The downside is added component cost, while these are often meant for batteries, which have steady voltage, more than capacitors, which have widely varying voltage.

Let's define $\Delta E$ as the energy quantum threshold that triggers one accumulation event or counter update. This defines the resolution of energy tracking. Meanwhile, $E_{v}$ is the energy consumed to process one counting event (i.e., detect that $\Delta E$ was reached and increment a counter or log the event). $P_{id}$ is the constant background (quiescent) consumed power. Assuming the measurement circuit is placed between the transducer and storage, the energy consumption can be expressed as
\begin{align}
    E_c(t_0,t)=(t-t_0)P_{id} + \Big\lfloor \frac{E_H(t_0,t)}{\Delta E}\Big\rfloor E_{v}.\label{count}
\end{align}
Because energy is integrated in discrete quanta of size $\Delta E$, the instantaneous energy state is only known with resolution $\Delta E$. Therefore, the measurement error here is dominated by this quantization process, i.e., $\epsilon\in[0, \Delta E)$. Assuming uniformly distributed error and no bias correction, one has $\mathbb{E}[\epsilon]=\Delta E/2$ and variance $\sigma^2=\Delta E^2/12$. Other possible measurement error sources appear, for instance, if the system is too slow to detect very fast energy surges, as events might be missed, leading to underestimation, or if the analog components used for accumulation, like charge pumps or integrators, are leaky, as the accumulated energy may decay and cause early or missed counts \cite{Sigrist.2017,Naderiparizi.2016}.

Finally, note that alternative placements for the measurement circuit are possible as indicated in Section~\ref{sec:MPo} and \ref{sec:acquisitionM}, e.g., after the storage, allowing measurement of the net available or consumed energy. In such cases, the accounting in \eqref{count} needs to reflect both energy inflow and outflow, and the counting behavior would follow the actual energy dynamics rather than only cumulative input.

\subsubsection{Indirect EI}

EH sensors may be deployed to gauge the ambient source rather than the electrical output. Examples include a light sensor for solar, a thermometer pair for thermal gradient, an accelerometer for vibration, or an RF detector diode for RF strength. Using these, the ZED gathers environmental data that correlates with energy availability, hence providing indirect ETO measurements. They may be passive or active, and their power consumption depend heavily on the specific sensor type as discussed later in Section~\ref{sec:sensing}. Indeed, simple passive sensor-based indicators may provide a basic binary indication of ambient energy availability, e.g., presence or absence of sufficient illumination, without detailed quantitative measurements. In general, the sensors may be duty-cycled for reduced energy overhead. Meanwhile, measurement errors are influenced by i) the sensor-internal measurement error due to thermal noise, quantization (in ADC-based systems), resolution limits, and temperature/power-induced drifts; and ii) sensor-to-harvester mapping error since the correlation between the sensed variable and the actual energy output of the harvester is nonlinear and context-dependent, e.g., temperature/angle-dependent in the case of solar panel output. Note that even small absolute errors may represent a significant portion of the available energy budget, e.g., as in \cite{Rajappa.2023}, wherein an absolute error around 1–2 mW is obtained in an indoor solar EH scenario. There might be a need to regularly perform a correlation analysis, locally or edge-assisted, to maintain the mapping functions' accuracy under varying conditions.

\begin{table*}[t!]
    \caption{Typical energy sources of EH-IoT systems and their key features and predictability}
    \label{tab:sources}
    \centering
    \begin{tabular}{p{1.65cm}|p{1.4cm}|p{2.9cm}|p{1.6cm}|p{9cm}}
    \toprule
    \textbf{energy source} & \textbf{magnitude} & \textbf{key features} & \textbf{predictability} & \textbf{typical energy predictors}  \\ \midrule
     light    & medium to high (tens mW) & long cycles (e.g., diurnal for outdoor, working periods for office indoors)  & high & time-series models (e.g., ARIMA, exponential smoothing); periodic models (e.g., sinusoidal regressors, daily/weekly patterns); linear regression or decision trees using time, light level, temperature as features;  lookup tables based on local sunrise/sunset and light sensor patterns; and lightweight LSTMs or temporal convolutional neural networks (NNs) if HW permits \\ \hdashline
     kinetic, mechanical &  low (a few mW) & event-driven, contextual, \& user-dependent (e.g., walking, running) & low, context-dependent & activity classification models inferring movement types from accelerometer; signal analysis (e.g., step/peak detection, zero-crossing, frequency estimation); shallow NNs or support vector machines for motion pattern detection; although often more effective to sense EH patterns directly as context cues than to predict \\ \hdashline
     thermal & low to medium (a few mW) & stability over short intervals, dependence on device workload \& ambient conditions & low-medium & linear models using temperature gradient history; piecewise models based on thermal transition states (e.g., when a machine is active); some TinyML approaches if patterns emerge from context (e.g., heating cycles in appliances or machines)\\ \hdashline
     (ambient) RF  & very low (nW$-\mu$W) & dependence on RF sources proximity, reflections, \& occupancy & very low & hard to predict directly due to the limited energy budget, thus often treated as opportunistic\\
     \bottomrule
    \end{tabular} 
    \begin{flushleft}
    \scriptsize{
    The characterization of the listed energy sources reflects typical behaviors/techniques, while large deviations are possible for specific use cases/applications. For instance, a piezoelectric transducer harvesting kinetic/mechanical energy from trains' activity over a roadway can harvest W-level power with high predictability \cite{Sanislav.2021}.}
     \end{flushleft}
\end{table*} 

Another indirect EI acquisition approach relies on SW to assist or even avoid HW measurements \cite{Eriksson.2023}. For instance, firmware can compute useful aggregates of raw HW voltage/current readings instead of relying on HW processing alone, realizing a SW Coulomb counter that can track EI in the long term at the cost of some careful timing. Meanwhile, without HW measurements, nodes may infer their energy availability via time-to-event monitoring. The core idea is to track the elapsed time between specific, well-defined system-level events that are causally linked to energy accumulation or depletion. These events could be voltage threshold crossings, task completions, energy harvester activations, or ZED brownouts. For voltage threshold crossing in capacitor-based systems, the intervals reflect either the rate of EH if the system is charging or energy usage if it is discharging, depending on system configuration and load conditions \cite{Islam.2020}. Meanwhile, in the case of time-to-discharge monitoring, the time between power-on and brownout reflects the energy budget and, indirectly, the energy cost of the operations performed during that window. This time data can be used for several purposes, such as estimating energy budgets without active measurement, inferring the average power profile of recent activity, deciding when to schedule or delay operations, or calibrating statistical energy models, as in \cite{Geissdoerfer.2022}. This approach offers a nearly free signal that can be used jointly with the previous EH acquisition methods and that, when interpreted correctly, can yield actionable insight into energy dynamics.

Finally, another indirect way to acquire EI is to look into the future. This is precisely the goal of forecasting energy availability and/or consumption patterns. This is facilitated by the previous EI acquisition methods via data provision, and constitutes the scope of the following discussions.\\

\begin{remark}
     Obtaining energy measurements inherently introduces energy consumption (even negating EH benefits if not implemented carefully), requires specialized components and techniques, and increases overall system complexity and costs. This calls for overhead-aware and ultra-low-power/cost HW techniques for minimum-overhead energy measurements.
\end{remark}

\subsection{Energy Forecasting}\label{sec:intelligence}
The edge intelligence paradigm is gaining increased traction due to the AI boom and refers to on-device or at the network edge (or both) intelligence. TinyML is commonly used to provide application intelligence locally (on-device) and/or to optimize application-layer tasks, such as sensor sampling, anomaly detection, or local inference in typical IoT setups \cite{Kallimani.2024,Lopez.2023}, and must account for energy availability fluctuations when incorporating EH processes. Meanwhile, network edge intelligence focuses on configuring transmission parameters, scheduling uplink opportunities, or managing computation offloading from constrained nodes \cite{Lopez.2023}.  Indeed, edge intelligence can help adapt/optimize the IoT system operation based on current or predicted harvested energy, but here we focus specifically on its use for energy forecasting. Discussions on energy-aware operation protocols/concepts are provided later in Section~\ref{sec:protocols}, and may expand to include intelligence-based optimization.

Different ambient energy sources exhibit distinct temporal dynamics, magnitudes, and environmental dependencies, which directly impact the design and effectiveness of energy prediction models in EH-IoT systems, as captured in Table~\ref{tab:sources} (cf. \cite{Smart.2016,Unlu.2018,Ma.2020,Akan.2018,Zeadally.2020,Sanislav.2021,Leemput.2023,Lopez.2023,Banotra.2023} for more detailed energy source characterizations). For instance, solar energy follows a highly predictable diurnal cycle, enabling accurate forecasting with periodic or regression models as illustrated in the following example.
\begin{example}[Solar Prediction using ARIMA~\cite{Rajappa.2023}]
\label{ex:arima_solar_prediction}
    Short-term energy forecasting in solar-powered EH-IoT systems may rely on an auto-regressive integrated moving average (ARIMA) model. In this approach, solar irradiance is forecasted using a univariate time series model trained offline on historical light sensor data. The resulting ARIMA model parameters, including autoregressive and moving average coefficients, can then be transferred to a low-power embedded system for use during operation.

    At runtime, irradiance measurements are collected via a light sensor and fed into the ARIMA model to predict near-future irradiance values at fixed intervals. These predictions are converted into expected harvested energy at time slot $n$, \( \hat{E}_{H}(n) \), using known parameters such as the solar panel area \( A_{pv} \) and efficiency \( \xi_{pv} \), PMU efficiency \( \xi_{pmu} \), and sampling period \( T \), which is also the time slot duration, according to 
    \begin{align}
    \label{eq:harvested_solar_energy}
    \hat{E}_H(n) &= \frac{1}{2}(\hat{I}_{r,0}(n) + \hat{I}_{r,1}(n))TA_{pv} \xi_{pv} \xi_{pmu},
    \end{align}
    where \( \hat{I}_{r,0}(n) \) and \( \hat{I}_{r,1}(n) \) represent the predicted solar irradiance at the beginning and end of time slot \( n \), respectively.
    
    One may now compute the waiting time required for a task with energy requirement $E_{task}$ to be safely executed as the minimum $N$, or $TN$, that satisfies $\sum_{n=1}^N \hat{E}_H(n)\ge E_{task}$. This predictive mechanism enables the device to remain in a low-power sleep state until sufficient energy is available, thereby optimizing energy use and reducing reliance on large energy storage.

    \begin{figure}
        \centering        \includegraphics[width=\linewidth]{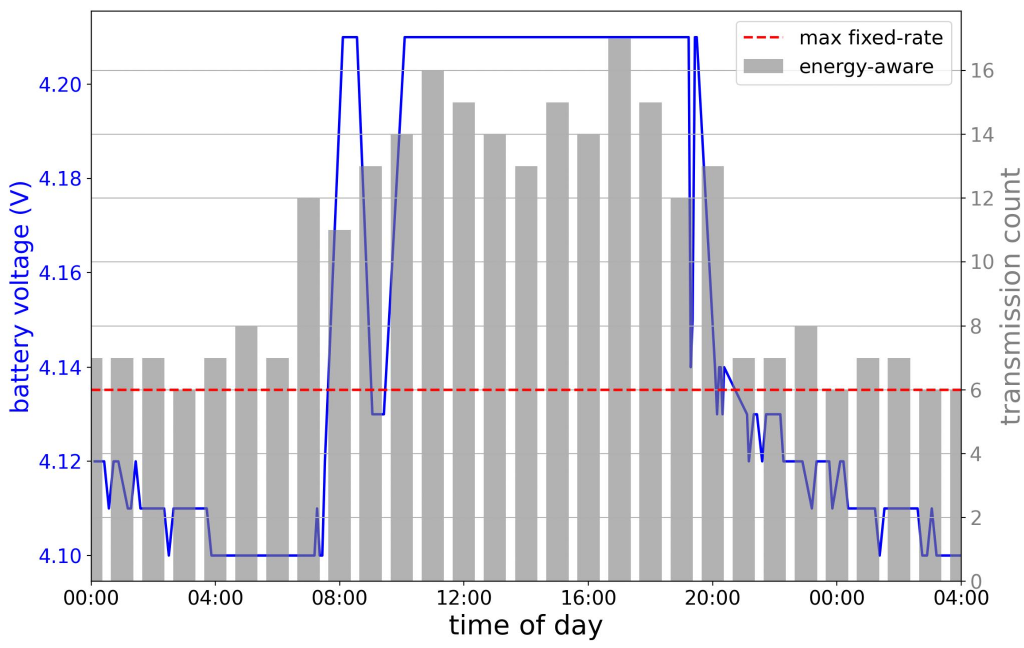}
        \caption{Battery voltage and number of sensor data transmissions of a solar-powered air quality logger with \( A_{pv} \) = 81x137 mm$^2$, $\xi_{pv}$ = 0.17, and $\xi_{pmu}$ = 0.85, as a function of the time of the day (12-05-2022) and using ARIMA.  The ARIMA model was trained on solar irradiance data collected over 120 days in Antwerp, Belgium, between August 2021 and February 2022 as ARIMA(5,1,0) with \(T\) = 30 s.  Herein, the autoregressive parameter, i.e., number of lagged observations, is 5, the differencing order is 1 for a stationary irradiance time series, and there is no moving average component.
        }        \label{fig:results_solar_prediction}
    \end{figure}
    
    The data transmission frequency of a solar-powered air quality logger using this method was evaluated in a field deployment. As shown in Fig.~\ref{fig:results_solar_prediction}, the logger achieved a significant increase in transmission frequency compared to a fixed-rate baseline while maintaining battery stability. This demonstrates the effectiveness of integrating statistical energy forecasting directly into embedded energy management logic.   
\end{example}
Kinetic and RF energy are highly event-driven and sporadic, requiring context-aware or reactive approaches. Indeed, the energy predictor choice depends on the source, and its complexity must be tailored to the ZED's CPU and energy budget \cite{Sandhu.2021,Famitafreshi.2021}, e.g., an IoT gateway or coordinator could run a sophisticated predictor using weather data while an 8-bit MCU might stick to a simple linear model or even a lookup table.

At a ZED, EI acquired at the ETO provides real-time and source-specific insight,\footnote{Some ETO-based EI datasets can be found in \cite{Smart.2016,Kuzman.2019,Sigrist.2019}, often including probabilistic models \cite{Smart.2016} and forecasting approaches \cite{Kuzman.2019}.} while EI acquired at the ESE offers a smoothed, integrated view of energy accumulation over time. The latter is suitable for estimating long-term energy budgets or planning duty cycles conservatively. All in all, the specific measurement point influences both the granularity and type of prediction model that can be used effectively.

Meanwhile, as mentioned earlier, ZEDs may get help from network edge nodes to reduce the sensing and computation burden from energy forecasting processes by offloading their prediction tasks and data. Moreover, network edge nodes may aggregate environmental data and/or share contextual insights that are otherwise too costly for individual ZEDs to acquire or process \cite{Jewsakul.2025}. For example, a gateway with access to weather forecasts, historical solar patterns, or local RF activity can perform higher-fidelity energy forecasting and transmit lightweight prediction summaries to nearby nodes, which share similar contexts. Edge network nodes may also fuse multi-node energy data to detect spatial energy trends (e.g., sunlit vs. shaded zones), enabling cooperative adaptation across a network. They can even host a digital twin of the ZEDs' energy, continuously updated with real-time and forecasted data (not only based on history but also indirectly by observing network observing communication activity patterns), as a central intelligence layer, allowing proactive energy management strategies and smarter decision-making that aligns with both individual ZED constraints and global network goals.

The network edge nodes may forecast ambient energy availability for a given area of interest. Indeed, they may share EH insights in the form of notification messages at the application level to subscribed ZEDs that are in, or plan to visit as part of their route, that area of interest. 
Such data (e.g., successful EH events experienced by a ZED in the past given its trajectory) can be crowdsourced by ZEDs registered to an EH forecasting service running at an edge host. This approach may extend the concept of predictive quality-of-service in an edge computing system deployment \cite{palaios2023} to a ZED context.
\begin{remark}
    On-device intelligence and network edge intelligence, either for energy forecasting/coordination or application support, should be seen as complementary rather than competing paradigms in EH-IoT scenarios. On-device intelligence enables low-latency decisions with minimum connectivity requirements, while edge intelligence offers broader context and coordination through aggregated data and more complex models. Hybrid approaches that balance both may enhance adaptability, efficiency, and scalability in energy-constrained environments.
\end{remark}
\section{Operation Tasks \label{sec:tasks}}
IoT devices perform tasks involving environment interaction (e.g., via sensing, actuation), local data processing, and wireless communication \cite{Ma.2020,Sandhu.2021,Lopez.2023}. This section presents a structured characterization of these tasks, focusing on their energy cost profiles, execution granularity, and responsiveness requirements. Such insight forms the basis for informed energy budgeting, scheduling, and adaptation in constrained environments.
\subsection{Sensing Tasks}\label{sec:sensing}
Sensing is the foundation of most IoT applications \cite{Banotra.2023,Hesam.2024}. Depending on the service-related task (e.g., recognition of an industrial machine's state, situational awareness enhancement for autonomous robot-host movement), sensors onboard ZEDs may measure temperature, pressure, movement, and others. Their energy cost can range from a few $\mu$J to hundreds of mJ, depending on the sensor type, sampling rate, interface complexity, and required pre-processing, i.e.,
\begin{itemize}
    \item low-power microelectromechanical systems (MEMS) sensors (e.g., temperature, humidity, light, accelerometers) typically consume $\mu$W$–$mW when active and can complete a measurement in 1$–$10 ms \cite{Zeadally.2020,Banotra.2023}, e.g., a digital temperature sensor may require from a few to tens $\mu$J per sample \cite{Sabovic.2020};
    \item barometers and magnetometers may draw hundreds of $\mu$W to a few mW, depending on resolution and sampling frequency \cite{Banotra.2023};
    \item gas, $\text{CO}_2$, and particulate matter sensors and spectrometers often require mW to W \cite{Majhi.2021,Banotra.2023}, long warm-up times in the order of minutes, and sustained current draw such that energy costs per reading can reach 100$–$500 mJ or more, exceeding typical harvested energy in a short time.
\end{itemize}
\begin{figure}[t!]
    \centering    \includegraphics[width=\linewidth]{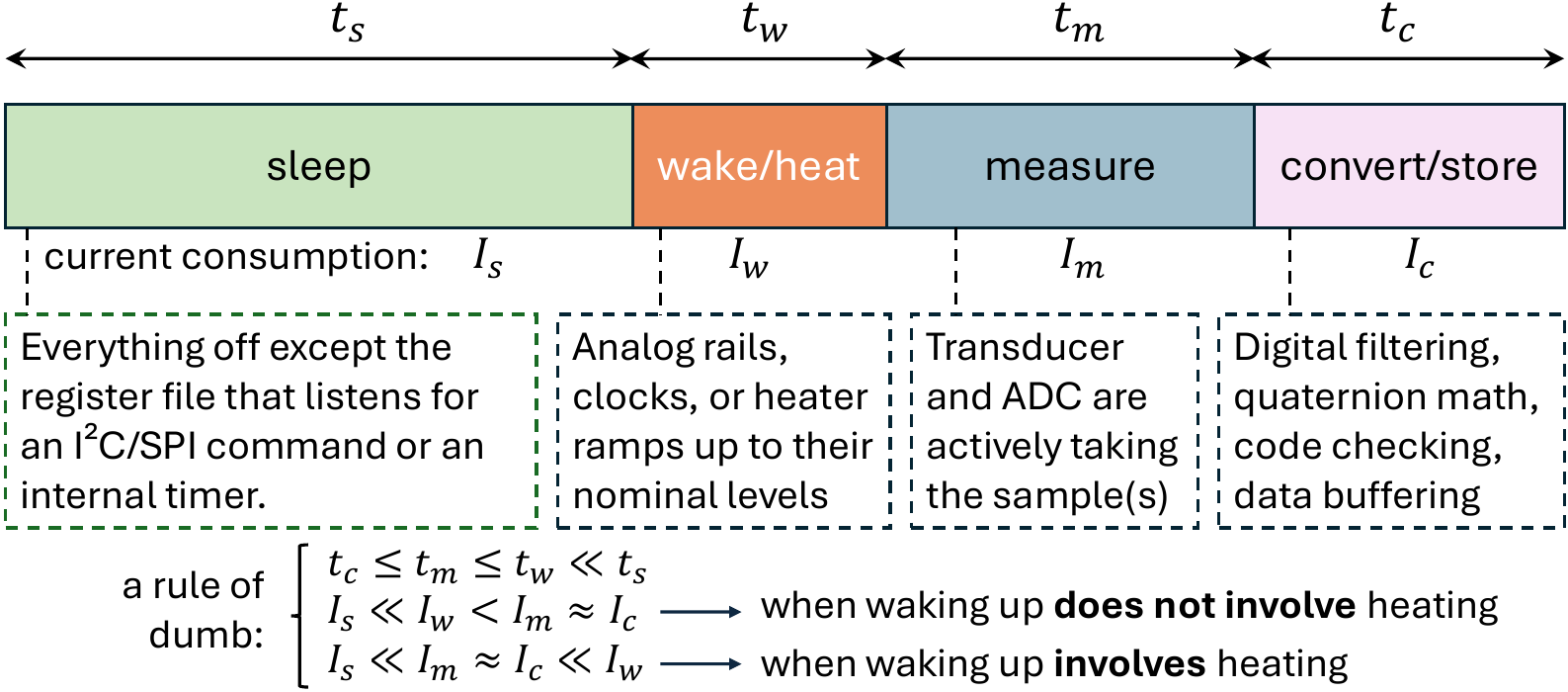}
    \caption{Typical sensor states in a full cycle operation. Note that heating during wake-up is typical in gas and chemical sensors, and a few biomedical and environmental transducers. These sensors rely on raising a material or active element to a stable elevated temperature before they can produce meaningful or accurate readings. This is a highly energy-consuming state.}
    \label{fig:sensorS}
\end{figure}
In general, a full-cycle sensing operation includes sleep, wake/heat, measure, and convert/store states, sequentially, as illustrated in Fig.~\ref{fig:sensorS}. They have different time and current consumption profiles, each state $i\in\{s,w,m,c\}$ consuming $E_i=V_{dd}I_i t_i$ energy units. Usually, $t_w,t_m,t_c$ are fixed, while $t_s$ may or may not be fixed, leading to periodic or on-demand sensing, respectively. Note that frequent short sampling is viable for sensors with low standby and startup energy, while increasing the sampling interval or storing more energy between cycles is needed for sensors with high warm-up or startup costs to sustain full operation. Some sensors, like accelerometer \href{https://www.analog.com/media/en/technical-documentation/data-sheets/adxl362.pdf}{ADXL362}, support onboard buffering or event-driven modes, triggering interrupts only when a change is detected and allowing the MCU to stay in deep sleep unless the sensor has meaningful data to report.

Finally, the interface between the sensor and the MCU can significantly influence energy use. Indeed, analog sensors require ADC conversions, adding conversion and processing cost, while communication overhead is added in the case of digital sensors (often faster and more precise) via $\text{I}^2\text{C}$ or SPI protocols \cite{Zidar.2023}. Notably, smart sensors with built-in processors can offload filtering, thresholding, or even inference, allowing the main MCU to remain asleep. In general, pre-processing at the sensor or early in the pipeline (e.g., filtering, downsampling) is energy-efficient when it reduces data transmission or defers wake-up of more power-hungry components.
\subsection{Computation Tasks}\label{sec:comp}
Computation is increasingly important as ZEDs shift toward more autonomous and intelligent behavior. Computation tasks have varying complexity, energy cost, and timing constraints, which may or may not be suitable for intermittent operation.

The computational capabilities of MCU-free devices are often minimal and highly specialized. Such devices rely on passive components or minimal digital logic, e.g., threshold detectors, amplifiers, and filters, to execute tasks such as identification, signal conditioning, or rudimentary logical operations. For example, passive RFID tags compute simple challenge-response protocols using HW-embedded cryptographic primitives, while devices incorporating physical unclonable functions (PUFs) carry out identity verification through the inherent randomness in their physical structure \cite{Sandomirskii.2025}.  Energy consumption in these cases is tightly bound to the amount of charge transferred through capacitive nodes or the duration of analog activity \cite{Hasler.2019}. However, precise modeling is often empirical due to the analog or mixed-signal nature of such implementations.

Meanwhile, devices embedding digital MCUs are more computationally versatile, capable of executing a wide range of programmable tasks. 
\subsubsection{Types of Programmable Tasks}
Programmable tasks are classified and
exemplified in Table~\ref{tab:Ncyc} and described in more detail next.

\emph{Basic logic, control, and low-level protocol management:}
This includes all event-driven logic and control flows such as finite state machines, conditional decisions, and reactive logic, e.g., evaluating whether to transmit data or trigger actuators \cite{Shnayder.2004}. These operations are typically lightweight and fast, consuming at most tens of $\mu$J per execution. Although very simple, they must be carefully managed in ultra-low-power devices to avoid unnecessary energy waste. Good practices include interrupt-driven execution, offloading repetitive timing to HW timers, and relying on deep sleep modes to minimize background power draw. This category also includes lightweight protocol-related computing, such as scheduling periodic transmissions, handling timers, and computing simple checksums or cyclic redundancy checks (CRCs).

 \begin{table}[t!]
    \centering   
     \caption{Examples of programmable tasks and the corresponding scaling of $N_{cyc}$ with the input size $n$}   \label{tab:Ncyc} 
    \begin{tabular}{p{2.4cm}|p{3.7cm}|p{1.6cm}}
    \toprule
    task type & specific tasks & cycle scaling \\ \midrule
      \multirow{3}{2.4cm}{basic logic, control, \& low-level protocol management}   & threshold check, timer interrupt, handling, scheduler tick & $\mathcal{O}(1)$ \\
      & CRC-8 update & $\mathcal{O}(n)$ \\
      \hdashline
       basic data processing and aggregation  & delta encoding, run-length encoding, min/max search& $\mathcal{O}(n)$\\
      & moving average (window $k$) & $\mathcal{O}(nk)$\\
      & simple histogram ($k$ bins) & $\mathcal{O}(n\!+\!k\log\! k)$\\
       \hdashline
      \multirow{3}{2.4cm}{signal processing and feature extraction} & FIR filtering ($k$ taps)  & $\mathcal{O}(nk)$  \\
      & FFT (radix-2) & $\mathcal{O}(n\log n)$ \\
      & autocorrelation & $\mathcal{O}(n^2)$ \\
       \hdashline
       TinyML inference & $K-$means update ($k$ clusters) & $\mathcal{O}(nk)$ \\
       & convolutional NN (kernel size $k$) & $\mathcal{O}(nk)$ \\
       & binary classifier & $\mathcal{O}(n^2)$\\
       \hdashline
       \multirow{3}{2.4cm}{security and cryptography operations} & HMAC generation & $\mathcal{O}(n)$ \\ 
       &ECC scalar multiplication  & $\mathcal{O}(n^2)$ \\
       & RSA encryption & $\mathcal{O}(n^3)$\\
       \bottomrule
    \end{tabular}   
\end{table}

\emph{Basic data processing and aggregation:}
This includes basic signal processing and data manipulation, such as moving averages, thresholding, min/max computation, or simple compression techniques like delta encoding or run-length encoding, which may reduce the communication/buffering burden.  For instance, computing a rolling average over 100 samples may take a few ms and tens of $\mu$J, while sending 100 raw samples over radio could cost hundreds of $\mu$J to mJ, i.e., $1-2$ orders of magnitude energy consumption difference. 
In-network aggregation, including summing, counting, or filtering data from multiple nodes, also falls here  \cite{So.2014}.  Note that while computationally cheap, such operations require maintaining local state and can accumulate substantial memory and energy cost over time in multi-hop or collaborative sensing networks.

\emph{Signal processing and feature extraction:} Tasks here involve mathematical transformations or data characterization that go beyond simple data processing/aggregation. This includes filtering (e.g., finite/infinite impulse response (FIR/IIR)), fast Fourier transform (FFT), convolution, correlation, and other statistical feature extraction like entropy and average energy. These operations are often required in acoustic, vibration, biomedical, or RF sensing applications, and their energy cost ranges in the order of $\mu$J to mJ depending on the task size and HW support. This category bridges the gap between low-level processing and full-scale inference, and benefits greatly from architecture-specific DSP libraries or HW accelerators. It is often a critical enabler of data reduction, relaxing the on-device data storage (or remote data offloading) requirements and real-time local analysis.

\emph{TinyML inference:}
As discussed in Section~\ref{sec:intelligence}, ZEDs may run lightweight ML models locally. They incur significant computational energy costs depending on the model size, memory footprint, data precision, optimization, and HW support. The usual approach is for models to be trained offline, e.g., at a network node, while running only the inference operations at the device. Still, small NNs such as keyword spotting or binary classifiers may consume tens to hundreds of $\mu$J per inference on optimized platforms, while this can rise to 1$–$10 mJ on general-purpose MCUs due to slower execution and lack of acceleration \cite{Ulkar.2021}. Prominent techniques to support the development or operation of TinyML models include \cite{Lopez.2023,Kallimani.2024}: architecture searching to find the best ML architecture fitting the available MCU resources, self-attention to weigh the input importance and enable parallelization, quantized or sparsified models to reduce memory and compute requirements, offloading pre-processing when possible, and triggering inference only when necessary, e.g., prefiltering with simple thresholds. Also, the required inference accuracy depends on the specific goal of the ZED. In some cases, exiting the TinyML model early during inference can save energy without compromising task fulfillment. As proposed in \cite{pomponi2024}, the idea is to pre-evaluate input samples and identify those where further layers are unlikely to yield significant accuracy gains.
Finally, note that despite local inference's high cost, it can eliminate costly communication overhead and is therefore increasingly central in network edge intelligence.

\emph{Security and cryptography operations:}
Security operations in low-power IoT are essential for data integrity and confidentiality and span three main buckets \cite{Grossschadl.2007,Kietzmann.2021,Kumar.2022}: bulk data protection, secure session setup, and lightweight integrity/policy checks.
Encryption and decryption are performed per message, with symmetric primitives like advanced encryption standard (AES) or hash-based message authentication code (HMAC) being often energy-affordable, especially with HW acceleration (e.g., consuming a few $\mu$J per block) \cite{Kietzmann.2021}. Meanwhile, public-key schemes like elliptic-curve cryptography (ECC) or Rivest–Shamir–Adleman (RSA) cost tens mJ and are reserved for infrequent events like device pairing or secure bootstrapping \cite{Grossschadl.2007,Kumar.2022}. Hence, good practices include offloading as much crypto as possible to HW engines, reducing on-device asymmetric operations, and adopting lightweight schemes.
Regarding authentication and key management, uninterrupted CPU and radio activity is required, which is challenging given the typical intermittency of many ZEDs. Finally, message integrity verification and access-policy enforcement incur only modest compute loads, tolerate deferred execution, and hinge purely on local context (e.g., sensor state, identity metadata).
\subsubsection{Processing Costs}
The energy consumed by any generic task on an MCU can be modeled as $E_{task}=P_{task} t_{task}$, where $P_{task}$ is the average power and $t_{task}$ is the execution time. The former is given by \cite{Rabaey.2002}
\begin{align}
    P_{task} = \underbrace{\gamma C_sV_{dd}^2f}_{\text{dynamic power}}+\underbrace{I_{leak}V_{dd}}_{\text{static power}}\ ,\label{Ptask}
\end{align}
where $\gamma$ is the activity factor representing the proportion of active gates, $f$ is the operating frequency, and $I_{leak}$ is the leakage current, usually significant only at higher temperatures or with sub-threshold voltage operation. Indeed, static power is usually negligible for CMOS circuits \cite{Liu.2012,Valente.2023}. Meanwhile, the execution time is given by 
\begin{align}
    t_{task}=N_{cyc}/f, \label{ttask}
\end{align}
where $N_{cyc}$  denotes the total number of clock cycles required. Using \eqref{Ptask} and \eqref{ttask}, the total energy consumption of a computation task can be written as
\begin{align}
    E_{task}= \underbrace{\gamma C_sV_{dd}^2N_{cyc}}_{\text{dynamic energy}}+\underbrace{I_{leak}V_{dd}N_{cyc}/f}_{\text{static energy}}\ ,\label{Etask}
\end{align}

Table~\ref{tab:Ncyc} exemplifies how $N_{cyc}$ scales as a function of the input size $n$ for several tasks belonging to the previously discussed computation categories. 
Note that even when tasks share the same asymptotic scaling law, their actual complexity or energy consumption can differ significantly due to differing constant factors. The latter depend, for instance, on the memory access patterns, algorithmic implementations, data precision, and HW-level support such as accelerators or instruction set optimizations.

Low-power computing systems may operate with fixed or dynamic voltage and frequency settings. Simpler MCUs found in low-cost/complexity devices typically operate at fixed voltage and frequency, lacking the HW infrastructure, such as integrated voltage regulators or PLLs, to support dynamic adjustment. These suit applications with static workloads, tight energy budgets, or stringent timing requirements. In contrast, more capable IoT devices often implement dynamic voltage and frequency scaling (DVFS) to adapt energy use in real time based on task demands. These systems include dedicated PMUs and clock domains, allowing them to optimize performance-per-Watt by increasing frequency and voltage during intensive computation and lowering them during idle or low-activity periods. This entails $f\propto V_{dd}$ in low-voltage regions due to CMOS gate delay characteristics \cite{Rabaey.2002,Liu.2012,Valente.2023}, while the approximate linear relationship only breaks at the extremes of the voltage range. 
Therefore,  $E_{task}$ in DVFS-capable IoT devices, and assuming negligible static energy consumption, can be written as
\begin{align}
    E_{task}\approx \gamma' f^2N_{cyc},\label{EtaskDVFS}
\end{align}
where $\gamma'$ is an MCU-dependent parameter, scaling with $\gamma C_s$.
Note that DVFS flexibility is especially valuable in systems with variable workloads and real-time constraints.

\subsubsection{Transient vs/and Intermittent Computing}
 Two fundamental computing paradigms for ZEDs are ``transient computing'' and `` intermittent computing''.\footnote{The terms are often used interchangeably, as in \cite{Min.2022}.}
The former focuses on lightweight computing techniques for the on-periods, such as stateless or crash-consistent routines, opportunistic execution, approximate computing, minimal memory footprint, and ML compression techniques (as those leading to TinyML implementations). Meanwhile, the latter applies to ZEDs with non-volatile memory/processors and focuses on efficient state preservation across power cycles to complete meaningful computation. 
This includes preventing data loss or corruption across power failures using checkpointing and state-retention methods \cite{Hester.2017,Maeng.2019,Maeng.2020,Ma.2020,Sumanth.2021,Min.2022,Karimi.2023}. 
These refer to inserting code snippets to periodically store the state of a running program, such that the system can recover from the latest stored state when power is restored, ensuring continuous progress.
\subsection{Communication Tasks}\label{sec:comm}
IoT devices include communication capabilities to report data to and/or receive operational instructions from the network. Herein, we focus on wireless communication capabilities due to their flexibility, ease of deployment, and suitability for energy-constrained, mobile, or hard-to-reach environments typical in EH-IoT scenarios.\footnote{Indeed, wired IoT devices can usually be powered directly from the grid, especially in industrial, building automation, or smart infrastructure settings, hence no EH capabilities needed.}

\begin{figure}
    \centering    \includegraphics[width=\linewidth]{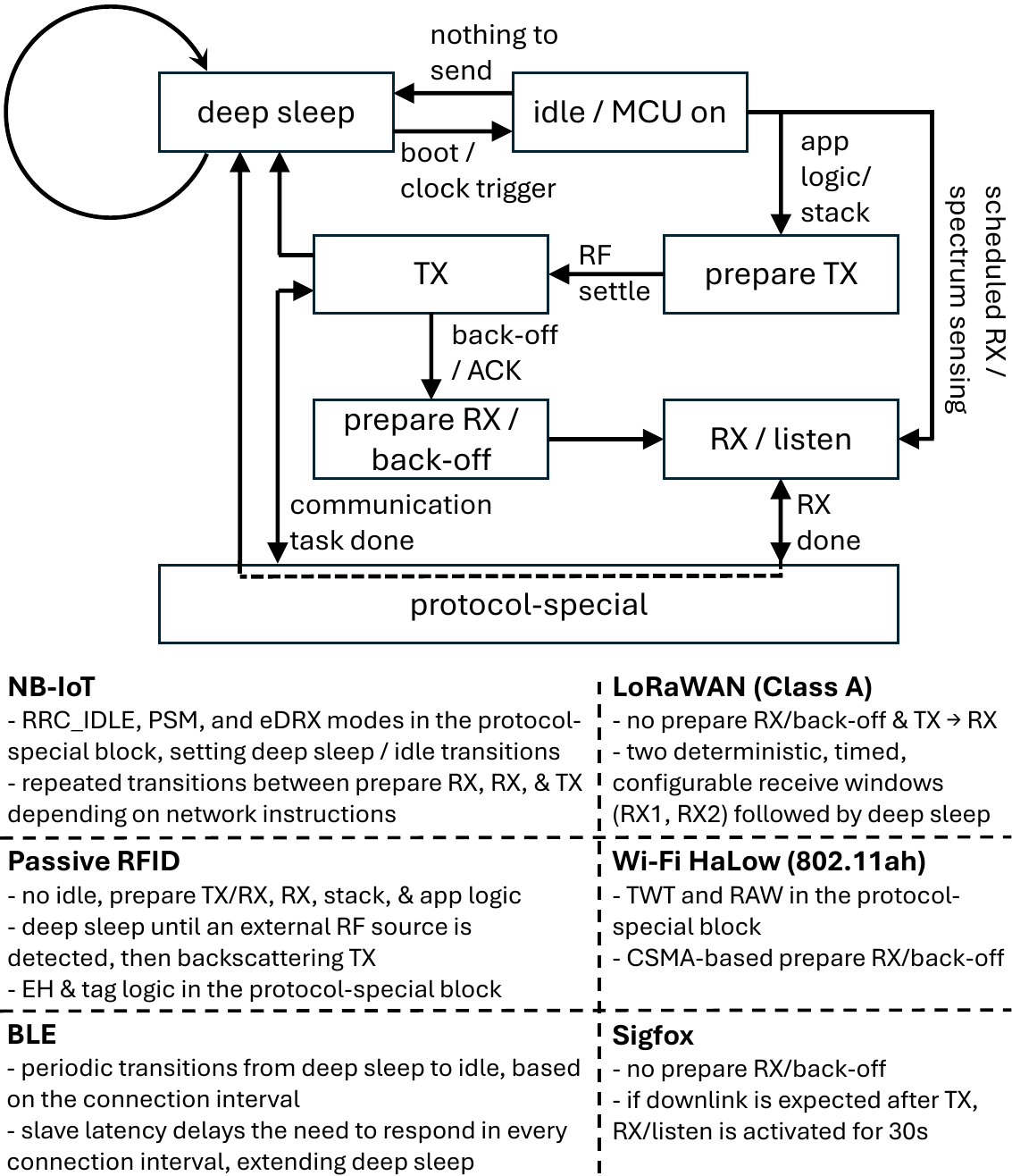}
    \begin{flushleft}
    \footnotesize{Acronyms: radio resource control (RRC), power saving mode (PSM), extended discontinuous RX (eDRX), target wake time (TWT), restricted access window (RAW), and carrier sense multiple access (CSMA).}    
    \end{flushleft}    
    \caption{(top) Canonical communication finite‑state machine for a low‑power IoT radio and (bottom) example of modifications for specific low-power IoT connectivity solutions.}
    \label{fig:commS}
\end{figure}

Fig.~\ref{fig:commS} depicts a canonical communication finite‑state machine for a low‑power IoT radio, tracing a tight loop through every energy‑relevant activity the node can experience. This applies to any generic low-power IoT connectivity protocol, although with specific variations as exemplified at the bottom of the figure. In general, power consumption is the lowest during deep sleep. Only a real‑time calendar or an external interrupt is alive therein, waking the MCU into an idle state to run application logic and packet preparation when needed. If active transmission (TX) is required, the radio prepares for it, e.g., energizing its phase-locked loop (PLL) and power amplifier bias, before entering TX mode. Immediately afterward, either the radio goes back to sleep or slips into prepare reception (RX) or back‑off if acknowledgements (ACKs) or contention are required, followed by an actual RX/listen state. In these RX states, the radio holds its low‑noise amplifier active just long enough to detect a preamble, beacon, or ACK. Once the MAC's guard conditions are satisfied (timeout, ACK received, schedule complete), the machine collapses back into deep sleep, ready for the next timer tick or external event. The time/energy profiles for each of the communication states depend greatly on HW characteristics, data rate, modulation scheme, and the underlying communication protocol specs.

Note that communication is often one of the most energy-intensive operations in EH-IoT systems. A key determinant of the actual energy cost is the communication modality employed, which can be i) active communication (as in bluetooth low energy (BLE), Sigfox, NB-IoT, Wi-Fi HaLow, and LoRaWAN), wherein the device's radio transceiver generates and transmits the signals consuming power in the order of several to hundreds of mW during TX/RX depending on technology/range \cite{Zeadally.2020,Leemput.2023,Banotra.2023}, and startup overheads (e.g., oscillator stabilization) further contributing; ii) semi-active/passive communication (as in backscattering with a logic powered by a local energy source), wherein the device modulates a (variable) data signal relying on an external carrier, entailing $\mu$W-range power consumption but limited range and data rate; and iii) passive communication (as in passive backscattering, and resonance-based tags), wherein the device modulates or reflects the external signal with purely passive components and an MCU is not often involved, leading to nW$-\mu$W power consumption \cite{Stanacevic.2020}.\footnote{An alternative classification is \cite{Ma.2020}: i) generative radio, which requires the transmitter to actively generate RF waves to carry information; and ii) reflective radio, which conveys information by (passively or semi-actively) modulating and reflecting RF waves impinging its surface.}

Synchronized local oscillators and digital processing blocks in typical active transceivers consume substantial energy that should be avoided for a large class of ZEDs. Indeed, active communication often consumes 1–2 orders of magnitude more energy than sensing or computation. 
Still, active TX/RX is feasible for ZEDs with a more relaxed energy budget if it is selectively triggered, batched, and carefully matched to EH and storage capabilities. Other techniques compatible with semi-active/passive transceivers such as non-coherent or analog-based RX methods, e.g., envelope detectors, energy detectors, or passive RF peak detectors, can operate at orders of magnitude lower power ($\mu$W or sub$-\mu$W) by detecting the presence of signal energy or simple modulation patterns without requiring full decoding \cite{Stanacevic.2020}. These receivers are ideal for wake-up signaling or simple binary communication where only the presence or absence of a pulse matters. The concept of packet-less transmission, wherein events' detection is conveyed by transmitting a single pulse (or bit) without using elaborate packet headers, is closely related to this and finds applications in structural health monitoring, chemical reaction detection, and other nano-scale IoT applications \cite{Ma.2020}. Note that these techniques trade energy efficiency for limited functionality as they often lack addressability, are prone to false triggers, and cannot support full data RX without additional circuitry. 

Meanwhile, adding confidentiality, integrity, and authentication on top of a lean radio stack, as required in many IoT applications, introduces its own energy and timing penalties \cite{Grossschadl.2007,Khan.2022}. Security routines entail not only computationally intensive crypto operations to protect packets, as mentioned in Section~\ref{sec:comp}, but also extra message exchanges to negotiate keys and memory reserved for session contexts, all of which keep the radio and processor awake longer. Indeed, security-related protocol interactions may impose strict timing constraints and state-transition challenges. For instance, integrity verification must coincide precisely with packet-reception events, forcing the radio and processor to remain active at exact instants. Also, multi-step key exchanges and session authentications depend on sequential message round-trips that are vulnerable to dropout and synchronization errors. More security-related discussions will follow in Section~\ref{sec:sec}. 

In general, ZEDs may benefit from aggregating data, minimizing transmission activity, avoiding idle listening, using lightweight/stateless protocols and piggyback control data onto payloads, and selecting communication schemes appropriate for the device's energy context \cite{Ma.2020,Zeadally.2020,Banotra.2023,Hesam.2024}.
\subsection{Actuation Tasks}\label{sec:act}
Actuation enables an IoT device to interact with or modify the environment, contrary to sensing, which focuses only on gathering information from the environment. Actuators are mainly continuous, e.g., DC motor for rotational speed control, or discrete, e.g., solenoid valve (on/off control), \cite{Banotra.2023}, and typically involve higher and burstier energy demands compared to sensing, communication, and low-complexity computation. Indeed, actuation energy consumption is not compatible with the energy budget of simple EH-IoT systems, and it has been much less explored/considered in the EH-IoT literature and industry for this reason. This is quickly changing with the advent of more efficient/powerful EH systems and low-complexity actuators, e.g., using ultra-low-power actuator drivers and shape-memory alloy-based actuators.

Actuators are typically driven through GPIOs, pulse width modulation (PWM) channels, or analog interfaces, often involving i) driver transistors or H-bridges for current amplification; ii) flyback diodes for inductive loads; and iii) capacitive energy buffering (e.g., supercapacitors) to handle high-power transients without collapsing system voltage. Moreover, actuation mechanisms vary widely in power and control complexity \cite{Adnan.2023,Banotra.2023}.

\begin{table*}
\caption{Typical power ($P$) and time ($t$) requirements for actuator classes common for low-power IoT devices and corresponding energy consumption computation example}
\label{tab:act}
\begin{tabular}{p{2.4cm}|p{1.3cm}|p{4.8cm}|p{3.5cm}|p{4.1cm}} \toprule
     \textbf{actuator class} & \textbf{$P$, $t$ req.} & \textbf{key energy-relevant parameters} & \textbf{energy consumption ($E$)} & \textbf{use example}  \\ \midrule
      MEMS micro-valve/mirror (e.g., for optical switching, $\mu$-fluidic routing) & 0.01-2mW, 20-500$\mu$s & $C_M\rightarrow$ electrode capacitance, $V_1/V_2$ pull-in/release voltage & $E=C_M(V_1^2-V_2^2)/2$ (if charge is simply dumped after switching, i.e., worst case) & Mirrorcle A3I8.2 dual-axis, gimbal-less MEMS mirror (0.8 mm diameter) with $C_M=100$ pF, $V_1=70$V, $V_2=0$V $\rightarrow E=0.25\mu$J\\
      \hdashline
      LED (e.g., status blink, heartbeat LED, IR) & 1-500mW, 0.05ms-1s & $V_F\rightarrow$ forward voltage, $I_F\rightarrow$ rated forward current, $t\rightarrow$ actuation time & $E=V_F I_F t$ & Kingbright WP7104 (red 3 mm) with $V_F=2$W, $I_F=10$mA, $t=20$ms 
      $\rightarrow E=0.4$ mJ \\ \hdashline     
      piezo bimorph/ stack (e.g., for micropump, haptic buzzer) & 0.05-2W, 0.05-5ms   & $C_p\rightarrow$ electrical capacitance of the piezo element, $V_d\rightarrow$ peak drive voltage
    & $E=C_p V_{d}^2/2$ (per actuation) & TDK PowerHap 1204 (part 1204H018V060) with $C_p\!=\!0.5\mu$F, $V\!=\!60$V 
    $\rightarrow E\!=\!0.9$mJ  \\ \hdashline
     latching solenoid, latching reed relay (e.g., valve poppet) & 0.05-1W, 5-100ms & $R_c\rightarrow$ coil resistance, $t_{1}$/$t_{2}\rightarrow$ operation/release pulse width, $V_{op}\rightarrow$ minimum/nominal operate voltage & $E=V_{op}^2t_{p}/R_c$ (per switch) with $t_{p}\in\{t_{1},t_{2}\}$ & Ledex 124-131-012 with $V_{op}=5V$, $R_c=160\Omega$, $t_p=20$ms, and push-pull, latching actuation $\rightarrow E=3.1$mJ per pulse\\ \hdashline
      e-ink panel (e.g., for sensor read-out label) & 0.05-0.5W, 0.1-1s 
      & $V_d\rightarrow$ logic/driver supply, $I_u\rightarrow$ update current, $t\rightarrow$ full-update time, $A\rightarrow$ panel area, $E_A\rightarrow$ energy-per-area figure & $E=V_dI_ut_u$ or $E=A E_A$ & 1.54" Waveshare e-paper module with $V_d\!=\!3.3V$, $I_u\!=\!18$mA, $t_u\!=\!0.5$s (full refresh) $\rightarrow E\!=\!30$mJ\\ \hdashline 
    shape-memory-alloy (SMA) wire/spring (e.g., for micro-latch, drug-delivery plunger) & 0.05-2W, 0.1-1s (+1-1.5s for passive cooling)  & $R_l\rightarrow$ electrical resistance, $I_{r}\rightarrow$ recommended current or current density, $\Delta T\rightarrow$ required temperature rise, $mc_p\rightarrow$ thermal mass & $E\!=\!I^2 R_l t_{heat}$ (per contraction), $t_{heat}\!\approx\! mc_p \Delta T/$ $(I_r^2 R_l)$ (with passive cooling afterward) & Flexinol\textregistered Wire $–$ 100 $\mu$m with $R_l=6.2\Omega$,  $I_r=0.25$A, $\Delta T=50$K, $mc_p=1.62\times 10^{-3}$ J/K $\rightarrow t_{heat}\approx 0.2$s, $E\approx 80$ mJ \\ \hdashline     
   micro-servo, miniature DC motor (e.g., for camera tilt, lock bolts) & 0.2-6W, 0.1-2s &  $V_s\rightarrow$ supply voltage, $I_0\rightarrow$ no-load current (free-run), $I_s\rightarrow$ stall current at zero speed, $K_t\rightarrow$ torque constant, $\omega_0\rightarrow$ no-load angular speed, $\theta\rightarrow$ rotation angle per move, $\tau_{L}\rightarrow$ external load torque during move, $I_{hl}\rightarrow$ current needed to hold position, $t_{hl}\rightarrow$ holding time & $E \!=\! E_{move}\!+\! E_{hold}$ (if used), $E_{hold}=V_s I_{hl}t_{hl}$,  $E_{move}\!=\!V_s I_{avg}t_{mv}$, where $I_{avg}\!=\!I_0\!+\!\tau_{L}/K_t$, $t_{mv}\!=\!\theta/(\omega_0(1\!-\!\lambda)$, $\lambda\!=\!\tau_L/(K_t I_{s})\Big)$
     & pan-tilt micro-servo (SG90-class) with $V_s=5$V, $K_t=0.29$ Nm/A, $I_s=0.6$A, $\omega_0=8.7$ rad/s, $I_0=0.04$A, $I_{hl}=0.15$A, $t_{hl}=1$s $\rightarrow I_{avg}=0.21$A, $t_{mv}=0.25$s, $E_{move}=0.26$J, $E_{hold}=0.75$J, $E\approx 1$J\\ \bottomrule
\end{tabular}
\begin{flushleft}
\footnotesize{This table has been compiled after processing datasheets from numerous manufacturers/providers, including  \href{https://www.bosch-sensortec.com/}{Bosch Sensortec}, \href{https://www.kingbrightusa.com/}{Kingbright}, \href{https://www.sensirion.com/}{Sensirion}, \href{https://www.mirrorcletech.com/wp/}{Mirrorcle Technologies}, \href{https://www.tdk-electronics.tdk.com/en}{TDK Electronics}, \href{https://www.ti.com/}{Texas Instruments}, \href{https://www.johnsonelectric.com/en}{Johnson Electric}, \href{https://www.waveshare.com/}{Waveshare}, \href{https://www.analog.com/}{Analog Devices}, \href{https://dynalloy.com/flexinol-actuator-wire/}{Dynalloy}, \href{https://protosupplies.com/}{Proto Supplies}, and \href{https://sps.honeywell.com/us/en/products/sensing-and-iot}{Honeywell Sensing and IoT}.}    
\end{flushleft}
\end{table*}

Table~\ref{tab:act} lists typical actuator classes for low-power IoT devices together with their power and time requirements and energy consumption computation framework. The latter is even exemplified based on actuators' datasheet. Note that:
\begin{itemize}
    \item LEDs, piezo buzzers, and MEMS mirrors/valves consume $\mu$W-mW, with pulse durations in the order of $\mu$s-ms;
    \item latching relays and miniature solenoids use a few mJ per switch, typically with sub-W peak power and pulses spanning from a few to tens ms;
    \item e-ink displays and SMA wires/springs draw tens to hundreds mJ per update, typically within hundreds of ms cycles;
    \item micro‑servos, DC gear‑motors, and micro‑pumps require hundreds of mJ per actuation, typically with hundreds of mW bursts lasting hundreds of ms.
\end{itemize}
Indeed, actuation energy per event can range from sub-$\mu$J (e.g., MEMS micro-valve/mirror) to a few J (e.g., micro-servo for camera tilt), with timing and reliability constraints that preclude speculative or partial execution.
Whenever possible, it is advisable to use latching/bistable actuators, which require energy only during state transitions and trigger actuation immediately after energy availability peaks.

\begin{remark}
The intended application of a ZED dictates its operational tasks and design (as long as feasible), leading to a spectrum of classes, from ultra-simple, sleepy, and passive responders to highly functional and dependable nodes. Crucially, these tasks differ significantly in energy/time demands and execution continuity, exhibiting highly heterogeneous progress dynamics that operation protocols must take into account. For example, a multi-packet data upload can be gradually completed over intermittent energy cycles, while a short actuation pulse, such as triggering a relay, must succeed in a single uninterrupted burst to avoid complete failure.
\end{remark}

\section{Energy-usage Models and Trade-offs}\label{sec:usage}

The amount and timing of energy available for IoT task execution are tightly coupled to the specific EH, energy storage architecture, and circuit-level interfaces in use. This section formalizes key energy evolution and usage models based on the timing and structure of energy consumption and EH, abstracting away from protocol-specific mechanisms. It also discusses the granularity at which energy constraints can be enforced and highlights how storage-centric limitations, such as leakage, efficiency losses, and capacity boundaries, shape the feasibility and performance of these models. All these aspects define the foundational energy behavior of EH-IoT systems and influence protocol design constraints.

\subsection{Energy State Evolution}\label{sec:evolution}
The evolution of the energy available at the ZED's storage element over time, $E(t)$, can be modeled by tracking the energy inflow from EH and the energy outflow due to consumption and leakage as
\begin{align}\label{EA}
    E(t)=E(t_0)+ \eta_1 E_H(t_0,t)-\frac{1}{\eta_2}E_L(t_0,t)-E_{leak}(t_0,t),
\end{align}
where $E(t_0)$ is the energy in the storage at a previous time $t_0$, while $E_H(t)$ is the harvested energy, $E_L(t)$ is the load energy consumption, and $E_{leak}(t)$ is leakage energy, all in the interval $[t_0,t]$. Moreover, $\eta_1$ is the fraction of harvested energy successfully stored as some is lost due to power conversion losses (e.g., from rectifiers or DC-to-DC converters \cite{Asadi.2022}), charging inefficiencies (e.g., internal storage resistance), and voltage mismatch; while $\eta_2$ is the fraction of stored energy effectively delivered to the load as some is lost due to discharge inefficiencies (e.g., voltage drop due to internal resistance), converter inefficiency (e.g., when boosting storage voltage to match the load requirements), and leakage paths or transient inefficiencies during discharge events \cite{Alves.2021}. 

\begin{figure*}[t!]
    \centering    \includegraphics[width=0.95\linewidth]{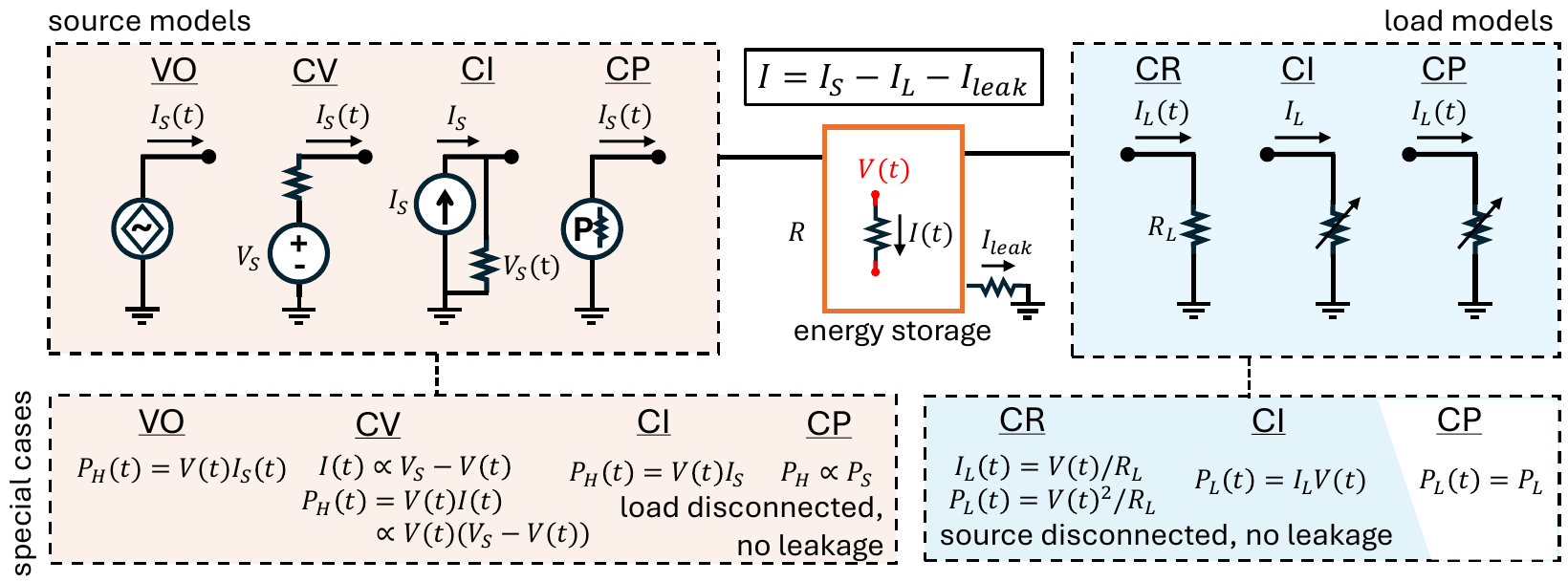}
    \caption{Source and canonical load models, and illustration of the computation of the harvested power and load power consumption for some special cases.}
    \label{fig:circuits}
\end{figure*}

Some modeling artifacts/constraints are often incorporated into \eqref{EA} when addressing optimization problems, either related to EH, storage, or load management, to ensure physics-law consistency. For instance, at any time, $E(t)$ cannot grow larger than the energy storage capacity, $E_M$, neither be negative, i.e., $0\le E(t)\le E_M, \forall t$.

\begin{remark}
    ZED’s future behavior is tightly linked to past EH and energy consumption, encouraging predictive and state-aware operation policies.
\end{remark}

The specific modeling of the harvesting, consumed, and leakage energy, and even the energy state evolution in \eqref{EA} depends on how the EH sources and load behave and the type of energy storage. Fig.~\ref{fig:circuits} summarizes the main source and load models, which are discussed in detail in the following, and how the harvested power and load power consumption look in special cases. The special cases are i) harvested power given disconnected load and no leakage, and ii) load power consumption given disconnected source and no leakage. In more general cases, where source, load, and leakage are simultaneously active, the net harvested current $I(t)$ (shown on top of the storage element in Fig.~\ref{fig:circuits}), is the one governing the battery state evolution through $E(t)=E(t_0)+\int_{t_0}^t I(\tau)V(\tau)d\tau$, wherein the charging voltage $V(t)$ evolution depends on the specific storage.
\subsubsection{EH source behavior}
The EH source and its interface circuitry may provide a i) variable output (VO), ii) constant voltage (CV), constant current (CI), or constant power (CP), ordered from low to high complexity.

\textbf{VO sources} provide variable voltage $V_S(t)$, current $I_S(t)$, and power $P_S(t)$, and are typical of EH circuits without transducer control, wherein it is not possible to enforce any of the other source behaviors discussed in the sequence.\footnote{Actual power harvesting can only occur at time $t$ if $V_S(t)>V(t)$.} A typical example is that of a solar cell without PMU, which exhibits a nonlinear I–V behavior that depends on the illumination (irradiance) and temperature \cite{Martynyuk.2019}.
    
\textbf{CV sources} maintain a fairly constant output voltage $V_S$, either naturally or with the help of voltage regulators \cite{Asadi.2022,Martynyuk.2019}. For instance, this may be the case of i) thermoelectric generators (TEGs), which output a voltage roughly proportional to the temperature gradient and do not fluctuate rapidly; ii) piezoelectric harvesters with rectifier and zener regulator, wherein a voltage clamping circuit makes the output stable; and iii) regulated solar harvesters with low-dropout regulators. Herein, as $V(t)$ increases, becoming closer to $V_S$, the harvesting energy decreases. This is why some EH systems disconnect or bypass the storage at high voltages and serve the load directly using PMUs or control logic. Maximum power is generally transferred when the load voltage is around half the source voltage, due to impedance matching \cite{Asadi.2022}.

\textbf{CI sources} deliver a roughly steady current $I_S$ as in the case of solar harvesters under stable illumination and operating below the maximum power point and some kinetic harvesters producing relatively stable current pulses per motion event \cite{Sabovic.2020,Delgado.2022,Martynyuk.2019}. In general, some active current regulation, e.g., current source ICs or feedback loops, is required.

\textbf{CP sources} maintain constant power delivery $P_S$ as in the case of EH circuits using PMUs with MPPT, which ensures operation at a voltage-current pair that delivers maximum power, and/or DC-to-DC converters \cite{Asadi.2022}. In general, continuous voltage and current monitoring, fast feedback, and dynamic adjustment of impedance or duty cycle are required. 

Note that CP operation always requires dynamic regulation, while CV, CI, and VO may be inherent to the source or achieved via simpler circuits.
\subsubsection{Load behavior}
A connected load may behave as a constant resistance (CR), constant current (CI), or constant power (CP) load \cite{Price.1993,Kalbitz.2020}, ordered from low to high complexity.

\textbf{CR load} with fixed resistance $R_L$ (but also generalizable to a fixed impedance \cite{Price.1993}) 
is the most passive and forgiving load model, wherein current and voltage vary proportionally.
Here, as the storage discharges and $V$ drops, the load’s power consumption drops quadratically. Such a load consumes disproportionately high power when the storage is well charged, while storage's energy 
slowly decreases toward the end, extending the device’s runtime at low power levels.  This load type might represent, for example, a sensor bias network, a heating element, an unregulated analog block, or any component whose current draw scales with supply voltage. 

\textbf{CI loads} draw the same current $I_L$ regardless of the storage voltage. This is typical for current-controlled actuation, like LED drivers or analog bias circuits. Notably, the power and energy consumption scales linearly with the storage voltage, instead of quadratically as in the CR case.  

\textbf{CP loads} draw a fixed power $P_L$ regardless of the supply voltage, and are common in regulated processors, radio modules, and digital circuits running at a fixed voltage.  The load dynamically adjusts its input current, with the downside that as voltage drops, current demand rises, possibly destabilizing the system or exceeding current limits near brownout.

\begin{remark}
In practice, the energy storage element of an IoT device may perceive the aggregate effect of different types of load components, which may be captured using polynomial, exponential, and more sophisticated weighting models, as described in \cite{Price.1993}.
\end{remark}
\subsubsection{Storage-related considerations}
The type of energy storage influences the charging/discharging time dynamics in the energy state evolution stated in \eqref{EA}. 

In the case of capacitors, the energy evolution dynamics can be explicitly derived from voltage and current using physical laws \cite{Kalbitz.2020,Leemput.2023,Lacerda.2024}, with closed-form expressions for several combinations of source and load models.\footnote{For instance, cf. \cite{Kalbitz.2020} for the cases of CV and CI sources and CP loads.} This is because the I-V and time relationship in a capacitor with charging voltage $V$, charging current $I$, and capacitance $C$ is well defined and given by
\begin{align}
I(t)=C \frac{dV(t)}{dt},\label{eq:IC}
\end{align}
while the stored energy obeys
\begin{align}
    E=\frac{1}{2} C V^2.\label{eq:EC}
\end{align}
Not all of this energy is usable, though, but only $E'=\frac{1}{2} C (V^2-V_0^2)$, where $V_0$ is the cut-off voltage \cite{Kalbitz.2020}.

Meanwhile, battery dynamics are more complex due to electrochemical and hysteretic effects that decouple terminal voltage from instantaneous current and stored charge. Voltage behavior depends on SoC, internal resistance, chemical kinetics, temperature, and aging, preventing closed-form expressions of energy $E(t)$ from voltage and current alone. Taming this complexity calls for empirical or semi-empirical models using nonlinear differential equations where SoC evolves with net current, accounting for rate-capacity effects, diffusion, and relaxation \cite{Galushkin.2020,Dicke.2020,Hermanns.2015,Petr.2018,Rakhmatov.2001}. For instance, Peukert’s law \cite{Galushkin.2020} models discharge-rate-dependent capacity loss in legacy chemistries like lead–acid and NiZn, but is less relevant for modern Li-ion cells. The kinetic battery model \cite{Dicke.2020} splits charge into immediately usable and bound components with inter-compartmental flow, capturing recovery and rate effects. It has been mostly applied to model sensor node lifetimes under pulsed loads \cite{Hermanns.2015}. Equivalent circuit models, commonly used for Li-ion and Li-polymer cells, emulate dynamic voltage response using SoC-dependent voltage sources and passive elements like resistors and capacitors, though mainly in high-power contexts, with few efforts addressing low-power systems, e.g., \cite{Petr.2018}. Lastly, the diffusion-based electrochemical model by Rakhmatov–Vrudhula \cite{Rakhmatov.2001} captures ion transport to estimate charge availability over time, particularly suited to embedded low-rate discharge scenarios. 

\begin{remark}
Selecting the appropriate battery model is crucial for modeling the energy dynamics of ZEDs with rechargeable batteries. It must balance fidelity and complexity based on the application’s energy scale, timing sensitivity, and computational resources.    
\end{remark}

\subsection{Harvest-Use Interaction Models}\label{sec:class}
\begin{figure*}
    \centering
    \includegraphics[width=0.99\linewidth]{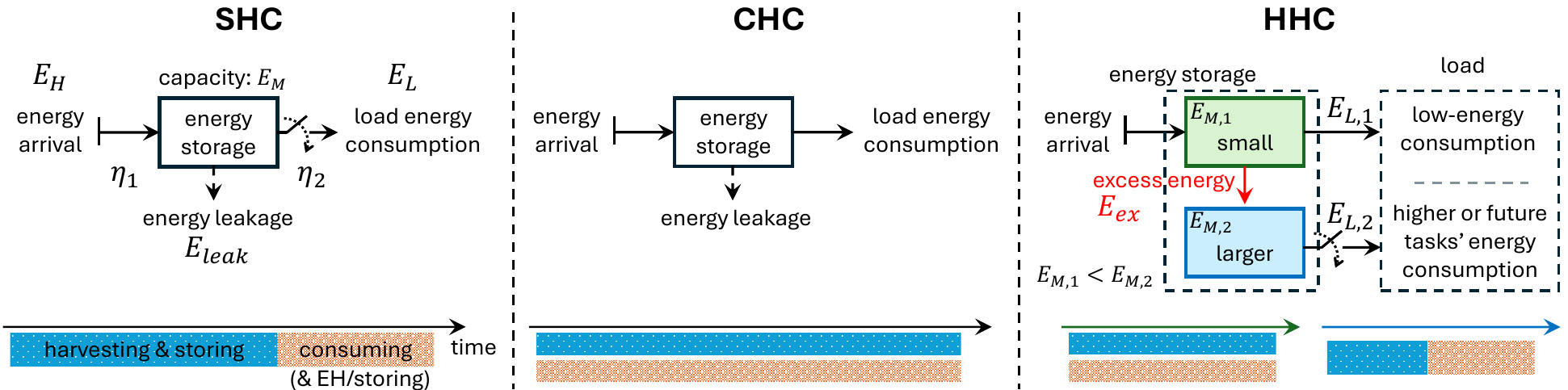}
    \caption{Energy-usage architectures and protocols, and corresponding stored energy evolution.}
    \label{fig:Eusage}
\end{figure*}

The relationship between harvesting, storage, and use motivates the following classification of energy-usage models, as illustrated in Fig.~\ref{fig:Eusage}:\footnote{This differs from the classification on harvest-use, harvest-store-use, and harvest-use-store, e.g., used in \cite{Lee.2015,Sanislav.2021,Alves.2021,Sandhu.2021}, in that such one focuses primarily on the order of energy storage usage and ignores potential simultaneities. Indeed, the harvest-use interaction modes in this paper focus more on the temporal relationship between EH and energy consumption, capturing runtime interaction and protocol implications more natively and enabling more nuanced analysis of system dynamics, scheduling, and efficiency trade-offs in real EH-IoT design.}

\subsubsection{Sequential harvest-then-consume (SHC)} energy is first harvested and stored, and consumption occurs only after sufficient energy accumulation \cite{Islam.2020}.  The load remains disconnected from the energy storage during the harvesting phase to support optimal impedance matching and MPPT without load interference, hence optimized EH efficiency. This mode is essential when dealing with extremely low or intermittent ambient energy sources since energy may often be insufficient to support simultaneous load operation. 

As SHC involves two phases, two energy state evolution equations are needed. One can use \eqref{EA} with $E_L(t)=0$, i.e., load disconnected, in the only-EH phase, and $E_H(t)=0$, i.e., harvester disconnected, in the load consumption phase. Note that the harvester may not need to be disconnected in the load consumption phase, but the energy harvested in such a period can still be ignored because of the typically small duration, but more importantly, because the EH efficiency drops significantly due to the reduced impedance matching.

\subsubsection{Concurrent harvest-and-consume (CHC)} energy is simultaneously harvested, stored, and consumed, allowing continuous or semi-continuous operation \cite{Islam.2020}. This is viable in environments providing relatively stable/moderate energy sources, and necessitates more sophisticated power-management HW, e.g., advanced PMUs or multi-stage converters to maintain voltage regulation and protect storage integrity under variable input/output conditions \cite{Lee.2015}. In this mode, \eqref{EA} can be applied as it is.
    
\subsubsection{Hybrid harvest-consume (HHC)} energy is initially buffered in a small storage element of capacity $E_{M,1}$ and immediately available for low-energy tasks, and the excess energy is transferred into another larger or more permanent storage element of capacity $E_{M,2}$ for sequential use. This combines concurrent and sequential aspects using multiple storage stages: immediate energy consumption from a small buffer and subsequent sequential use from a larger one. This model supports both SHC and CHC as special cases by reconfiguring energy paths or operating conditions dynamically, e.g., CHC to be used in high EH conditions while switching to SHC when EH becomes unstable/low. 

HHC provides further flexibility in energy use management at the cost of extra energy storage, hence complexity/cost. As an example, consider scenarios involving infrequent but energy-intensive tasks such as TinyML model maintenance or fine-tuning, secure firmware updates, or burst-mode data transmissions. When the energy cost of such operations is known or predictable, e.g., based on model size, update payload, or prior transmission behavior, this energy can be budgeted for the larger storage element in advance. This allows high-energy tasks to execute reliably without compromising the availability of the smaller buffer that supports routine tasks such as sensing, event monitoring, or beaconing.

As there are two storage elements in HHC, their state evolution must be assessed separately (although with input/output connection links). For the small storage, one can use the CHC energy state evolution modeling by substituting  $E_L(t)$ by $E_{L,1}(t)$, and $E_{leak}(t)$ by $E_{leak}(t)+E_{ex}(t)$, where $E_{ex}(t)$ is the excess power from the small energy storage that goes into the larger energy storage at time $t$. Meanwhile, for the larger storage, one can use the SHC energy state evolution modeling, involving two equations, by substituting  $E_L(t)$ by $E_{L,2}(t)$, and $E_H(t)$ by $E_{ex}(t)$.

\begin{remark}
Infrequent cycling in SHC may limit responsiveness in highly dynamic environments, while frequent cycling may accelerate storage degradation. Meanwhile, high energy leakages can reduce overall energy availability in CHC or HHC setups. 
Moreover, energy charging and consumption rates are quite different for most embedded ZEDs, with the former significantly lower \cite{Colin.2018}. This leads to negligible charging during operation and charging times orders of magnitude longer than discharge times. In such cases, the energy buffer is the only relevant power source during active operation, and its size determines the tasks that can be executed. 
\end{remark}

Each of these modes constrains the granularity, frequency, and intensity of task execution, and directly impacts protocol timing, energy availability, and task prioritization logic.

\subsection{Energy-usage Granularity}\label{sec:gran}
\begin{figure}[t!]
    \centering
    \includegraphics[width=\linewidth]{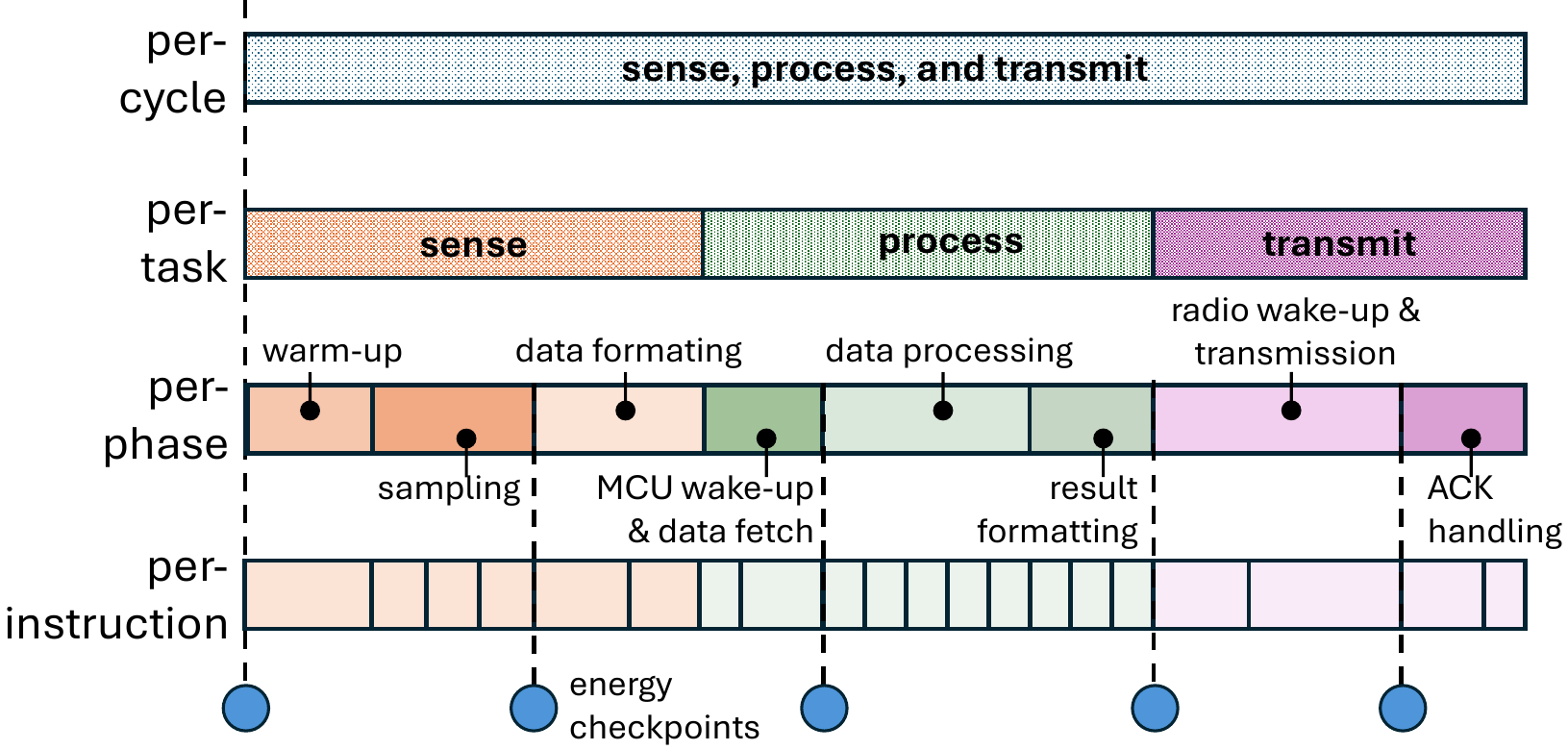}
    \caption{Example of energy-usage granularity: per-cycle, per-task, per-phase, and per-instruction.}
    \label{fig:granular}
\end{figure}

How energy constraints are enforced across operational tasks can vary significantly in EH-IoT systems. The main possibilities are illustrated in Fig.~\ref{fig:granular} and discussed next. A specific choice influences the precision required for EI acquisition and the system's flexibility, including operation protocols. 

Per-cycle (or per-chain \cite{Karimi.2023}) energy enforcement is the coarsest level. Herein, an entire operational cycle, typically comprising several related tasks, is only initiated if a conservative estimate of the total energy requirement can be met. This model is simple to implement and requires minimal runtime energy tracking. However, it may lead to inefficient use of available energy, particularly when waiting to accumulate sufficient energy for a full cycle introduces unnecessary latency or results in unused energy leakage. A more refined alternative is per-task energy budgeting, where energy usage is monitored and controlled individually for each task. This requires either offline profiling or online estimation of task-level energy cost and enables more responsive and opportunistic execution. Also, it supports dynamic task reordering and prioritization but depends on accurate and timely EI to avoid brownouts. Interestingly, tasks/cycles may be broadly grouped into three groups \cite{Colin.2018}: i) intermittent tasks/cycles with low energy demands that can tolerate being paused and resumed opportunistically, thus mainly computational tasks; ii) capacity-constrained tasks/cycles, which require a minimum stored energy to execute atomically, such as sending a 25-byte Bluetooth packet (35 ms at high power) or sampling from a sensor (8 ms at low power); and iii) temporally-constrained tasks/cycles, which are reactive and require energy to be available on-demand, such as transmitting in response to an external event. It is worth mentioning that certain processes, which can be seen as cycles or tasks, cannot be separated further. That is the case, for instance, of authentication and key management, which must run without interruption during pairing and involve computing and communication aspects (cf. Section~\ref{sec:comm}).

Further down in granularity, per-phase energy control partitions tasks into distinct functional phases, each with its energy profile. 
\footnote{Refer to Section~\ref{sec:tasks} for discussions on the different phases/states per tasks.}
This allows a system to scale down or omit certain phases based on real-time energy state, improving adaptability but demanding finer energy tracking and tighter HW/SW coordination. 
Finally, per-instruction-level energy tracking involves checkpointing and restoration of computational state based on instantaneous energy states in some intermittent computing platforms \cite{Ma.2020}. This offers maximum flexibility and resilience to power loss at the cost of added control overhead, and is discussed in more detail in Section~\ref{sec:Ecomp} on energy-aware computing.

Note that computational tasks often have more per-phase/instruction granularity, allowing significantly more subtask splitting compared to sensing, actuation, and TX. Independent of the granularity, EH-IoT operations must be designed as atomic execution units \cite{Karimi.2023,Delgado.2022}. An atomic task is one that must either be completed entirely or not executed at all. These tasks are typically made idempotent, meaning they can safely be retried after a power failure without side effects or corruption, and are energy-bounded, ensuring they can complete within the expected energy availability window. 
For single-energy-buffer ZEDs, the buffer must be provisioned at design time to support the largest atomic task in the system, while multi/reconfigurable buffers offer extra flexibility \cite{Colin.2018}. 

\begin{remark}
    The design of energy-aware systems must align energy-usage granularity with atomic task structure. 
\end{remark}

In coarse-grained models (e.g., per-cycle or per-task), each atomic task serves as the basic scheduling unit, while finer-grained models (e.g., per-phase or per-instruction) require additional mechanisms, such as checkpointing or energy-aware task partitioning, to enforce atomicity or emulate it dynamically  \cite{Ma.2020}. Task classification may be crucial before scheduling decisions \cite{Karimi.2023}.
\subsection{Storage-centric Trade-offs and Constraints}
Appropriate selection and implementation of energy usage models depend on the characteristics of the energy storage element. We list key energy storage technologies in Table~\ref{tab:energyS} and corresponding features, while more details and specific sub-categories can be found in \cite{Zeadally.2020,Calautit.2021,Banotra.2023}. 

\begin{table*}[]
    \centering
    \caption{Rechargeable energy storage technologies compatible with EH and key features}
    \begin{tabular}{p{1.6cm}|p{4cm}|p{1.6cm}|p{0.6cm}p{0.6cm}|p{2.2cm}|p{1.8cm}|p{2.3cm}}    
    \toprule
    \textbf{storage type} & \textbf{description} & \textbf{capacity} & \textbf{$\eta_1$} & \textbf{$\eta_2$} & 
    \textbf{energy leakage}
    & \textbf{life cycle}$^*$ & \textbf{common use case}  \\
    \midrule
      capacitor   & small-value ceramic or electrolytic capacitor (nF$-\mu$F range) & 1 $\mu$J $-$ 10 mJ &
      $\approx 1$ & $\approx 1$ & $0.5-5\%$ /hour 
      & $>10^6+$ & basic buffering, short bursts\\ \hline
      supercapacitor 
      & high-capacity (mF$-$F) capacitors with fast charge/discharge & 0.1 J $-$ 100 J & $0.85-0.95$ & $0.85-0.95$ & $0.5-5\%$ /hour
      & $>5\times 10^5$& intermediate energy buffering \\ \hline
      Li-ion / Li-Po battery & rechargeable battery with high energy density and decent efficiency & $0.5$ J $-$ 10 kJ & $0.80-0.90$ &	$0.85-0.95$ & 
      $\le 0.005\%$ /hour
      & $3\times 10^2-10^3$ & long-term storage, moderate loads \\ \hline
      solid-state battery & thin-film or printed battery with excellent leakage performance & 1 J $-$ 10 kJ & $0.75-0.90$	& $0.80-0.90$ & 
      $\le 0.001\%$ /hour
      & $10^2-10^3$ & wearables, medical, ultra-low power\\ \hline
      hybrid capacitor	& supercapacitor + battery chemistry (e.g., Li-ion capacitor) & 0.5 J $-$ 1 kJ & $0.85-0.95$ & $0.85-0.95$ &  $0.01 - 0.2\%$ /hour
      & $10^4-10^5$ & balance power density and life \\
     \bottomrule
    \end{tabular}
    \begin{flushleft}
    $^*$ Given by number of cycles, each comprising one full charge followed by one full discharge. Still, some storage types, like batteries, degrade slower under a shallow depth of discharge or partial cycling, which is common in EH systems and can extend effective lifetime significantly if managed properly.    
    \end{flushleft}    
    \label{tab:energyS}
\end{table*}

The energy capacity limitations of the storage element determine the size and type of tasks that can be executed without external intervention. In small-capacitance systems, medium-to-high power tasks such as RF transmission may not be feasible under a single charge cycle, requiring either energy aggregation over time (using SHC) or storage reconfiguration mechanisms. Conversely, oversized storage experiences longer charge-up times, delaying execution and reducing system responsiveness.

Another critical constraint is storage efficiency, comprising input ($\eta_1$) and output ($\eta_2$) factors \cite{Alves.2021}. Notably, low-harvest-rate systems charging a battery may face poor energy conversion efficiency due to high internal impedance or low charge acceptance rates, particularly under cold temperatures or aging. These losses imply that not all harvested energy is practically usable, and energy usage protocols must account for this discrepancy, either by incorporating safety margins in energy estimation or by dynamically adjusting task execution to avoid overconsumption.

Meanwhile, energy leakage is particularly relevant in (super)capacitor-based systems \cite{Alves.2021}. Capacitors exhibit non-negligible self-discharge rates, often on the order of $\mu$A, which can lead to substantial energy loss if energy is accumulated but not promptly used. As a result, sequential models like SHC may suffer from efficiency degradation when the storage dwell time is long, leaning the scale in favor of concurrent or hybrid models that promote immediate or periodic energy consumption. Indeed, the latter may reduce idle charge retention and improve overall energy utilization.

Finally, storage degradation over time, e.g., due to battery aging, capacitance drift, or equivalent series resistance increase \cite{Alves.2021,Calautit.2021}, impacts the suitability and predictability of energy storage technologies and energy usage models. This is not a concern for capacitors and supercapacitors, which are highly durable and ideal for frequent charge/discharge cycles, but it might be for the other technologies. Note that a protocol that relies on a fixed voltage threshold for triggering execution may become increasingly inaccurate unless it adapts to changes in actual energy availability caused by such degradation. This further motivates the inclusion of runtime energy estimation and possibly adaptive recalibration mechanisms to maintain robustness.
\section{Energy-aware Protocols \label{sec:protocols}}
This section focuses on operation protocols that use EI to coordinate and adapt task execution in EH-IoT systems. Recall that EI may refer to any knowledge about the EH process, stored energy, and/or load energy consumption, obtained either through direct or indirect measurements, including forecasting processes. Interestingly, while considering the corresponding EI acquisition overhead is paramount, as indicated in Section~\ref{sec:acquisitionM}, it is often overlooked in the literature. Indeed,  research works with a theoretical component incorporating or analyzing the EI acquisition overhead within their energy-aware protocol designs are scarce, as captured by Table~\ref{tab:refs}. Again, modeling such EI acquisition overhead deserves careful attention to advance the field. 

Next, we examine how protocols may leverage energy dynamics to govern sensing, communication, computation, and actuation procedures, emphasizing practical strategies, system-level mechanisms, and key challenges. Note that security aspects, previously falling within the computing and communication tasks umbrella in Section~\ref{sec:tasks}, receive here separate attention for convenience.

\begin{table*}[t!]
\caption{Key research on energy-aware protocols in the period 2018-2025 explicitly considering EI acquisition overhead}
\label{tab:refs}
\begin{tabular}{p{0.6cm}|p{2.4cm}|p{1.9cm}|p{2.65cm}|p{2.85cm}|p{2.9cm}|p{2.7cm}}
\toprule
    \textbf{ref} & \textbf{setup} & \textbf{EH \& storage} &  \textbf{EI \& acquisition HW} & \textbf{EI used in}  & \textbf{EI acquisition overhead} & \textbf{key results}  \\
     \midrule
     \cite{Islam.2020} & generic ZED with MCU running periodic and reactive tasks, each with timing and energy constraints & generic EH source (but solar is tested) \& generic storage & storage voltage using MCU's ADC and high-value resistor; energy requirements/gains of computational/EH tasks are profiled & deciding whether to execute computational tasks or harvest energy; and for adjusting scheduling decisions dynamically & considered in scheduler (time overhead is $1.2\%$ ms, energy overhead is not quantified); EI is acquired always before executing a task &   all admitted jobs are timely completed, outperforming baseline schedulers even with $12\times$ overhead  \\ \hdashline     
      \cite{Maeng.2020} &  generic ZED with MCU & generic EH source (but RF is tested) \& capacitor & capacitor voltage using MCU’s ADC and dynamic profiling of energy usage by measuring voltage drop during task/event execution & local scheduling of time-critical events (periodic or reactive), estimating the feasibility of scheduled workloads, and triggering quality-of-service degradation & all considered, while the proposals minimize and absorb their costs, e.g., memorizing measurement degradation levels to avoid frequent measurements & reliable periodic execution even with 10\% RF power fluctuation; significantly superior to benchmarks, specially in scheduling reactive events   \\ \hdashline 
      \cite{Sabovic.2020} & LoRaWAN ZED (implemented with SODAQ ExpLoRer board) modified to run from a capacitor and periodically measuring/reporting temperature & generic EH source with fixed EH rate \& capacitor & capacitor voltage and built-in ADC & local sensing/TX scheduling based on an EI threshold & offline measurements for sensing (17 mA, 8 ms), TX (50 mA, 50 ms), listening for ack (17$–$28 mA), deep sleep (0.4 $\mu$A), and power-off to power-on cycles, which require rejoining the network (13 mA, 13 s), and this is incorporated into the scheduling model & 
 the protocol avoids attempting too frequent updates in low-energy conditions (which would waste energy on repeated reboots), finding the sweet spot where the total energy cost per delivered reading is minimized \\ \hdashline
 \cite{Lopez.2022} & ambient RF-EH multi-antenna ZEDs & RF \& generic storage  & generic ETO and related HW & configuration of the antenna phase shifters for RF energy combining & time and energy consumed by the EI acquisition and phase shifting circuitry are simulated  &  dynamic RF combining can outperform static configurations given well managed overhead\\ \hdashline
\cite{Geissdoerfer.2022} & BLE ZED without EI acquisition HW & generic (but solar/piezoelectric EH \& a multi- layer ceramic capacitor are tested) & no HW, instead, the times the ZED turns on/off are recorded to locally and on-line build a statistical model of the recharge interval & connection protocol for reliable bi-directional communication between two battery-free ZEDs by synchronizing their active periods & protocol overhead in terms of energy ($<7.1$ $\mu$J) and time (tens of $\mu$s to a few ms) consumption is quantified in real HW & 10$–80\times$ higher throughput than state-of-the-art, evincing that estimating EI improves communication efficiency \\ \hdashline
\cite{Jewsakul.2025} & star topology LoRa network wherein a cloud server process the data and sends feedback + testbed  & solar \& supercapacitor or rechargeable battery & generic local storage level and telemetry to cloud (nodes report energy status or model parameters occasionally) & probabilistic TX based on local energy prediction; cloud uses network-wide EI for clustering and resource allocation feedback & protocol overhead was modeled, while energy used for on-board TinML vs radio for remote inference was measured and compared in testbed experiments
    & $2\times$ network lifetime and $8\times$ less network-wide overhead; sharing EI with low overhead improves workload distribution among nodes
 \\      \bottomrule
\end{tabular}
\end{table*}

\subsection{Energy-aware Sensing}
Energy-aware sensing includes techniques that modulate sensing frequency or sensor modality based on available energy. Simple approaches are rule-based duty cycling, e.g., sleep longer and sample less often if energy is low, and dynamic voltage and frequency scaling to accommodate sensing (but also other operations, especially computing and actuation) to the energy budget \cite{Lopez.2023,Maeng.2020,Sandhu.2021}.
Meanwhile, more intelligent/fine-grained adaptive sampling is gaining attention recently, e.g., leveraging TinyML energy predictors \cite{Karimi.2022}. In the case of multi-modal sensors, different sensing functions draw different amounts of energy due to their underlying physical principles, sampling needs, and operational requirements. For instance, a gas sensor may require mWs to heat a sensing element, while a temperature or humidity sensor typically consumes only $\mu$Ws, and a MEMS accelerometer falls somewhere in between depending on the sampling rate (cf. Section~\ref{sec:sensing}). Therefore, different sensing modalities should be activated/scheduled based on available energy and application needs, including priorities/hierarchies.

Another promising direction is context sensing from EH patterns since the amount of harvested energy often reflects environmental or user context \cite{Ma.2020,Sandhu.2021,Lopez.2023}. For example, kinetic-powered wearable ZEDs can infer step counts from the distinctive signal peaks produced by the transducer during footstrikes, while thermoelectric harvesters can detect surface temperature changes through variations in harvested voltage/current (cf. \cite{Ma.2020,Sandhu.2021} and references therein for more examples). By replacing dedicated sensors with lightweight context inference algorithms based on EH signals, ZEDs can significantly reduce their energy consumption as long as the fine-grained EI acquisition cost is managed carefully. Two main approaches exist \cite{Ma.2020,Lopez.2023}: (i) analyzing the instantaneous EH signal, which facilitates context detection but requires frequent sampling; and (ii) analyzing the accumulated energy over longer intervals, which is more energy-efficient but yields coarser contextual resolution. In both cases, simple TinyML techniques can be used to extract useful context with minimal processing overhead.
\subsection{Energy-aware Computing}\label{sec:Ecomp}
Computational fidelity or complexity may adapt to energy availability by tuning for lower/higher precision arithmetic, deferring complex processing, and dynamically selecting between full inference and lightweight heuristics, e.g., early exiting inference whenever the model is confident enough using only a fraction of model layers for inference, thus leading to energy-aware transient computing. This is crucial since becoming operational after a turn-off can require substantial (charging) time for most embedded ZEDs, as discussed at the end of Section~\ref{sec:class}.

\begin{example} [Energy-aware TinyML Selection \cite{sabovic.2025}] \label{ex:tinyML_model_selection}
Let us focus on ZEDs with TinyML inference. Instead of the usual approach of deploying a single, heavily compressed (and typically low-accuracy) TinyML model to accommodate worst-case energy conditions, multiple TinyML models may be deployed, out of which one may be adaptively selected/activated based on real-time energy availability. The decision can be guided by the energy storage evolution dynamics. For instance, for the case of a batteryless ZED with a constant current source, one can use \eqref{EA}--\eqref{eq:EC} together with Kirchhoff's current law as shown on top of Fig.~\ref{fig:circuits} to write the expected post-inference capacitor voltage \( V' \) as
\begin{align}
\label{eq:capacitor_voltage_dynamics}
 &
V' = I R_{L} 
\left(1 - \exp\left( \frac{-T_{L}}{R_L C} \right)\right) + V \exp\left( 
\frac{-T_L}{R_L C} 
\right).
\end{align}
Herein, \( I \) is the incoming harvesting current, \( R_L \) is the equivalent resistance of the inference task, which is a function of the corresponding current consumption, \( T_L \) is the execution time, and \( V \) is the pre-inference voltage. These values can all be obtained through prior measurements and acquisition methods described in Section~\ref{sec:acqusition}, where we consider \(I\) to be constant in between sampling intervals. By evaluating \( V' \) for each candidate TinyML model, the system determines whether a model is feasible, i.e., whether it leaves the capacitor voltage above the minimum threshold required for operation. Then, the scheduler may select the most accurate model that satisfies this condition for each task to be executed.

An experimental prototype powered by a solar panel demonstrates this adaptive behavior. Two different convolutional NNs are deployed on the device to perform gesture recognition based on inertial measurement unit data. Both models are based on a larger original model, which is scaled down in size and complexity to different degrees. Both scaled models are then compressed even further using quantization and LiteRT~\cite{litert.2025} conversion, resulting in a small TinyML (STML) model taking up 6.1 kB of memory and a larger TinyML (LTML) model taking up 104.8 kB of memory. Due to their difference in size and complexity, they also differ in terms of energy consumption, with STML and LTML consuming 0.12 mJ and 1.46 mJ, respectively.

When energy is plentiful, the system selects LTML, while it automatically falls back to STML under low-energy conditions.
Fig.~\ref{fig:tinyml_model_selection} illustrates how the capacitor voltage evolves over time as the system switches between STML and LTML, indicated by gray diamonds and black squares, respectively. Furthermore, we can see a comparison of the number of correctly and incorrectly executed inference paths across different energy levels, which highlights the accuracy trade-offs between the STML and LTML models.

This adaptive example approach enables energy-neutral operation while maximizing inference utility in environments with fluctuating energy availability.

\begin{figure}[t!]
    \centering
    \includegraphics[width=\linewidth]{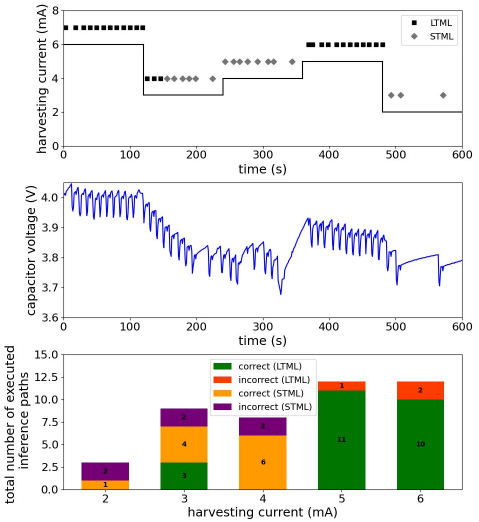}
    \caption{Capacitor voltage behavior over time when executing different TinyML inference paths with varying harvesting current for \( C = 0.5\,\text{F} \), with LTML and STML representing large and small TinyML inference for gesture recognition consuming 1.46 mJ and 0.12 mJ, respectively.}    \label{fig:tinyml_model_selection}
\end{figure}
\end{example}

While the above strategies (such as tuning for lower/higher precision models/arithmetic, and dynamically selecting between full inference and lightweight heuristics) are suitable for high-end ZEDs (i.e., ZEDs with reasonable computational capabilities), the strategies are limited and completely different in low-end ZEDs (i.e., ZEDs with limited computational capabilities). Essentially, a low-end ZED may only need to decide whether to execute or defer the arrived task based on the (actual or estimated) available energy, as exemplified next.

\begin{figure}[t]	
\includegraphics[width=\linewidth]{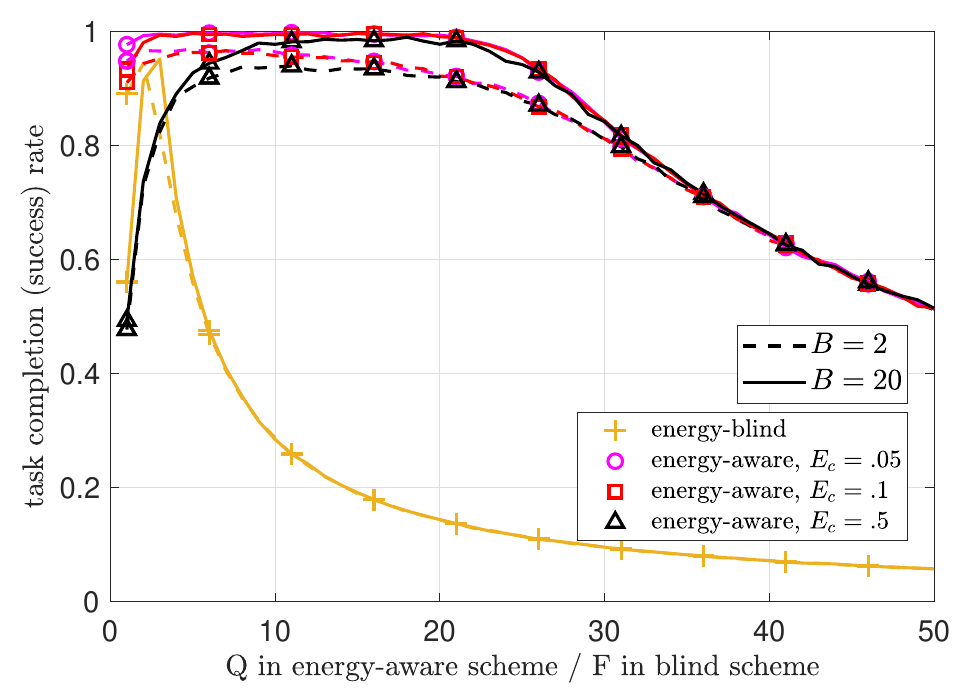}\\	\includegraphics[width=\linewidth]{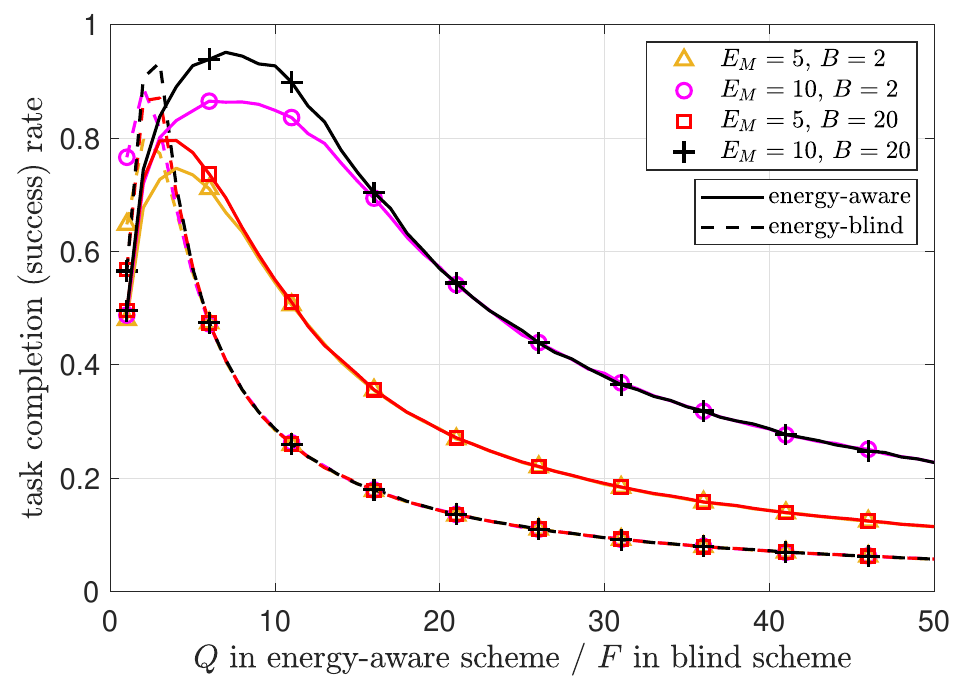}\\
\caption{Task completion (success) rate for the energy-blind and energy-aware schemes and different values of (top) $B$ and $E_c$, and (bottom) $B$ and $E_M$. We assume that the energy and task arrival processes follow a Poisson distribution with a mean of $0.75$ and a Bernoulli distribution with mean $0.35$, respectively. Furthermore, the task's energy consumption is set to $2$ units, and the total number of time slots to $10^5$.}
\label{fig:energy_meas}
\end{figure}

\begin{example}[Energy-aware Task Deferring]\label{ex:Mateen} 
    Consider a ZED with an energy storage capacity $E_M$, a computation buffer that can store up to $B$ tasks, and a time slot-based operation. In each time slot, the ZED can harvest energy and/or perform a computation task. 
    Without EI, the ZED may try to execute a buffered task at every $F$ time intervals. This is referred to as the "energy-blind" scheme. However, since the stored energy may not be sufficient to perform a task, an energy-aware mechanism for deferring its execution seems more appealing. For this, we let the ZED use $E_c$ energy units to measure the energy available in its storage every $Q$ time intervals. This EI is then used for the subsequent task execution decisions.
    
    In Fig.~\ref{fig:energy_meas}, we compare the task completion (success) rate for both schemes with different system parameters; that is, the portion of the tasks that are executed correctly given an infinite time horizon. We observe that the performance curves of both schemes exhibit concave-like behavior with respect to their periodic parameters (i.e., $Q$ and $F$ for energy-aware and energy-blind schemes, respectively). For the energy-aware scheme, a smaller $Q$ leads to a higher energy consumption for the measurement process and lower energy availability for the actual task execution. In contrast, the deviation between the energy estimate and actual energy becomes large for higher values of $Q$, contributing to wrong decisions regarding task execution. Overall, these effects produce a concave-like behavior for the energy-aware scheme's task completion rate curve. For the energy-blind scheme, the ZED opts to execute tasks more frequently when $F$ is small. In this case, any wrong decision has a twofold effect: i) an increase in the number of failed task executions, and ii) energy wasted in vain and affecting future task executions. For higher values of $F$, on the other hand, since the ZED waits longer to execute the task, the task buffer remains full with higher probability, and any arriving task will be dropped. Thus, leading to a smaller task completion rate. Combining these effects of small and large $F$ produces a concave-like behavior in the performance curve for the energy-blind scheme.
    
    As shown in the top figure, the successful execution of the tasks under the energy-aware scheme depends on $E_c$ for a given value of $Q$. Specifically, a higher $E_c$ results in performance degradation when $Q$ is small. Moreover, the size of the task buffer positively affects the performance of the energy-aware scheme, while the performance of the energy-blind scheme remains unaffected by the task buffer's size for larger values of $F$ and decreases with $B$ for smaller values of $F$. This is because a higher $F$ means that the tasks are executed after longer time intervals, increasing the chances that the task buffer becomes full and subsequently arriving tasks are dropped, thus, contributing to the task failure rate. On the other hand, for smaller values of $F$, a higher $B$ results in performance degradation because it is more likely that there are tasks in the buffer while the probability of having sufficient energy for successful task execution is low.

    From the bottom figure, and as expected, a higher $E_M$ is always beneficial. However, the potential gains are negligible for higher $F$ and $Q$ as the performance depends more on the task buffer size and the arrival rates in these regimes. In particular, if the task buffer becomes full, subsequent task arrivals will contribute to task failure even if enough energy is available. Obviously this also depends on the task arrival rate and energy arrival rate, but this requires further exploration, especially on how to optimize the system deployment and/or operation design parameters (such as $B$, $E_M$, $Q$ and $F$) for given operating conditions (such as task and energy arrival rates and $E_c$).

    Finally, note that the energy-aware scheme relies on conservative energy availability estimates, which may suit low-end ZEDs. Alternatively, a more sophisticated approach may leverage EH rate statistics while accounting for potential estimation errors and non-stationarity conditions.
\end{example}

\begin{remark}
Energy-aware intermittent computing is crucial under fluctuating energy availability to prevent ZEDs from attempting operations they cannot complete and thus avoid energy waste.
\end{remark}

Although previous practices can mitigate energy outages, the latter cannot be fully avoided, and ZEDs must resume their activity quickly and reliably whenever possible \cite{Ma.2020,Sandhu.2021,Lopez.2023}. Note that checkpointing consumes energy and time, thus, ZEDs must carefully decide when to checkpoint. Just-in-time checkpointing mitigates this issue by making checkpoints dynamically only when the power goes low \cite{Maeng.2019,Maeng.2020}. However, this becomes difficult for non-divisible operations, such as sensor access and flash writing, as they cannot be paused and resumed at an arbitrary point. This relates to our previous discussions in Section~\ref{sec:gran}, highlighting the need for atomic operation executions, and calls for more advanced scheduling strategies, potentially relying on energy availability forecasting. Moreover, it is appealing to insert potential checkpoints at compile-time and select which to trigger at run-time~\cite{Ma.2020,Lopez.2023}. Placing checkpoints after function calls, rather than inside them, reduces the number of variables saved, but spacing them too far apart may lead to ``Sisyphus loops'', where progress is repeatedly lost. This motivates watchdog-triggered checkpointing, which can forcibly break up long-running code paths; and energy-aware memory placement, which helps avoid state inconsistency, e.g., by allocating variables to volatile or non-volatile memory based on task idempotency and rollback behavior. Moreover, energy-aware timekeeping is crucial when ZEDs must synchronize sensing events or resume with accurate timing, calling for energy-frugal time-tracking mechanisms or energy-conditioned network time retrieval \cite{Lopez.2023}. All in all, a key challenge ahead is generalizing current advances for more complex applications and tighter real-time constraints. Indeed, ensuring correct execution and timing in unpredictable energy conditions becomes increasingly difficult as tasks get more complex, e.g., running AI inference intermittently.  
\subsection{Energy-aware Communication}
ZEDs must communicate opportunistically and with minimal overhead \cite{Lopez.2025,Yang.2019,Sabovic.2020}. Letting ZEDs advertise or transmit only when energy thresholds are met is an often effective, obvious strategy, but energy-aware protocols can go beyond this basic functionality. Indeed, the configuration of PHY parameters, such as modulation schemes, transmission power, and packet size, can adapt dynamically based on EI. For example, when energy reserves are low or channel conditions are poor, a ZED might switch to a more robust modulation scheme with/or lower throughput to reduce retransmissions, or reduce transmission power to conserve energy at the cost of shorter range. Similarly, transmitting fewer but larger packets can amortize fixed radio startup costs, improving energy efficiency in stable conditions. However, these adaptations require careful handling. For example, robust modulation schemes may be suitable at the TX side, but lower-complexity modulations are generally preferred to reduce RX energy consumption. Also, larger packets are more susceptible to corruption in noisy environments, potentially leading to costly retransmissions. Therefore, energy-aware PHY adaptation must carefully balance trade-offs between reliability, throughput, latency, and energy overhead, ideally guided by real-time measurements or predictions of channel quality and energy availability.

These principles can be concretely illustrated in real-world batteryless systems that integrate energy-aware communication and hardware-level adaptations. Indeed, we next exemplify how careful calibration of transmission intervals and power-saving features enables reliable operation under severe energy constraints.

\begin{example}[Energy-aware batteryless NB-IoT \cite{asultania.2023}]
Consider a large-scale ZED indoor deployment, such as smart home use cases or a solar-powered asset tracker, where wired power is infeasible and battery-based maintenance is both costly and environmentally unsustainable. NB-IoT, with its deep indoor coverage and low power profile, constitutes a promising connectivity option, and hence we  adopt the batteryless NB-IoT ZEDs powered solely by ambient indoor light using a small photovoltaic panel. Herein, the harvested energy is buffered in a capacitor, and the system operates based on dual voltage thresholds. Specifically, the ZED powers ON when the capacitor voltage exceeds $V_{\text{on}} = 4$ V and powers OFF below $V_{0} = 3.60$ V as shown in Fig.~\ref{fig:evoltage_thresholds}.  The PMU ensures that the capacitor voltage remains within safe and operable bounds, supplying 3.3 V to the NB-IoT chipset. The ZED starts communicating data packets until its voltage reaches the turn-off voltage. This threshold-based gating naturally aligns the ZED's operation with harvested energy availability, enabling transmissions only when sufficient energy is buffered. As a result, the ZED exhibits an emergent duty-cycled behavior where transmission frequency and uptime are implicitly regulated by real-time energy conditions.

\begin{figure}[t]\includegraphics[width=\linewidth]{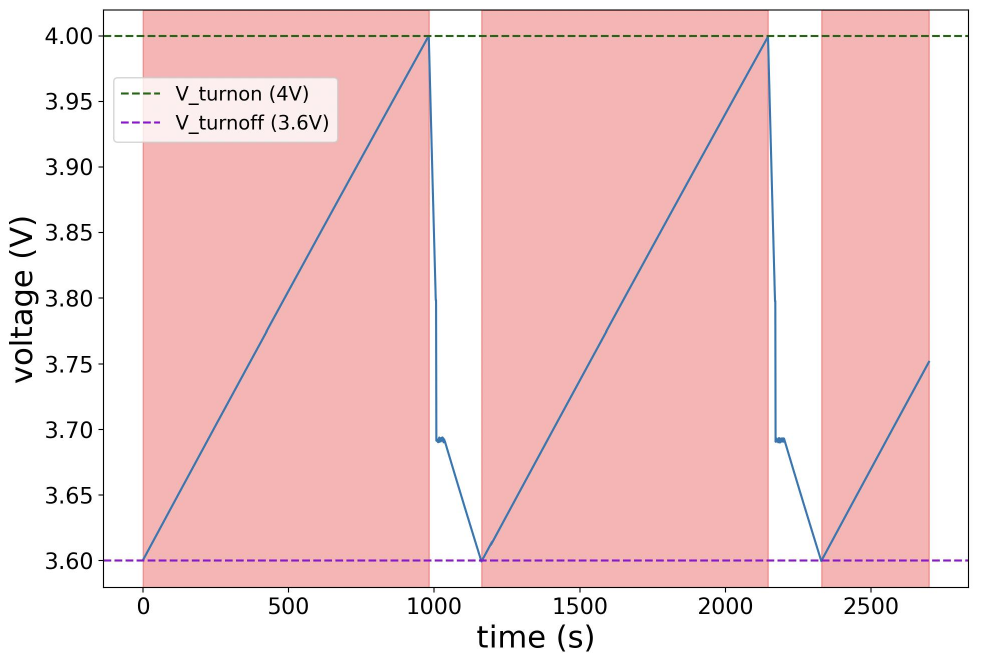}
\caption{The variation in the voltage of a 2.5~F capacitor during EH at 6~mW, with the ZED communicating unidirectionally.
}
\label{fig:evoltage_thresholds}
\end{figure}

The system’s energy-aware communication protocol accounts for the constraints of intermittent energy availability with the inclusion of dual voltage thresholds. Communication modes include i) unidirectional, where data packets are periodically transmitted in the uplink only; and ii) bidirectional, where every uplink packet expects a downlink response. The feasibility of these operations critically depends on three intertwined parameters: capacitor size $C$, transmission interval (TI), and average harvested power $P_H$. Experimental results demonstrate that for reliable transmission at $TI = 60$ s, $P_H \geq 6$ mW is sufficient to prevent restarts across all tested capacitor sizes. With $P_H = 50$ mW, even a small 1.5 F capacitor can sustain uninterrupted communication at $TI = 1$ s, achieving up to 3580 packets/hour in unidirectional mode and 2843 packets/hour in bidirectional mode. However, under reduced $P_H = 4$ mW, average packet throughput drops drastically to 90 (unidirectional) and 39 (bidirectional) packets/hour, underscoring the energy-aware system’s sensitivity to environmental lighting.

Moreover, enabling NB-IoT’s PSM after data transmission significantly reduces energy drain during idle periods, allowing the system to accumulate enough energy for subsequent transmissions. This is especially critical in natural-light scenarios, where energy availability varies with time of day and weather. Notably, there is a non-linear relationship between harvesting power and restart frequency. As $P_H$ increases, the ZED performs more frequent transmissions, consuming more energy and potentially increasing the restart count unless the capacitor is adequately sized. The optimal operating point, minimizing restarts while maximizing throughput, is experimentally observed at $P_H = 30$ mW for $TI = 1$ s. These results open up several promising directions for further refinement, such as adapting TI dynamically based on real-time capacitor voltage or predicted harvesting trends, and coordinating multiple ZEDs to share network resources efficiently under variable energy conditions. Integrating such mechanisms would further enhance the resilience, scalability, and sustainability of batteryless NB-IoT deployments.
\end{example}

MAC techniques for ZEDs, with emphasis on energy consumption reduction rather than energy-awareness provision and exploitation, are surveyed in \cite{Famitafreshi.2021}. In general, the MAC mechanisms and protocols vary according to the spectrum access regimes as discussed next. 

In shared-spectrum environments, such as cognitive radio networks and licensed shared access settings, energy-aware MAC design must accommodate the need for spectrum sensing and protection of incumbent transmissions. Here, unlicensed ZEDs may determine spectrum availability through energy detection techniques, assessing RF energy levels to identify transmission windows, termed ``white spaces'', for opportunistic access. This process involves non-trivial energy and timing overhead, as sensing, decision-making, and potential data transmission must be carefully balanced using harvested energy. Recent works \cite{liu2020cooperative,olawole2019cooperative} have proposed frameworks for optimizing the duration of sensing and energy allocation across tasks to maximize throughput under such constraints. In the case of RF-EH-powered ZEDs, EH patterns themselves can provide implicit context, where high energy influx may signal a busy channel due to primary transmissions, prompting nodes to defer access and vice versa. These tightly coupled energy- and spectrum-aware strategies highlight the dual role of energy sensing in both EH and medium access decisions.

Meanwhile, in non-(licensed-unlicensed) sharing scenarios, traditionally synchronized duty-cycling protocols are ill-suited for intermittently powered nodes, as any idle listening or coordination delay risks the node losing power mid-protocol. Grant-free uplink multiple access schemes, wherein nodes transmit opportunistically without prior scheduling or handshaking, are often appealing due to their simplicity and low coordination overhead \cite{Famitafreshi.2021,Lopez.2023,Sarang.2023}. This reduces idle listening and avoids synchronization costs, though at the risk of collisions. Still, note that grant-free MAC might be outperformed by either random or fast-uplink access protocols depending on energy budgets and traffic correlation \cite{Moons.2024}, thus, the optimal approach is scenario dependent. Meanwhile, downlink data transmissions may need to be triggered by the receiver, i.e., ZEDs, to avoid synchronization requirements \cite{SultaniaFamaey.2022}. Indeed, BLE has a receiver-initiated protocol feature called ``Friend Nodes'' that allows the devices to request buffered downlink data on demand, thereby reducing the need for constant synchronization and enabling energy-efficient communication.

A core energy-aware MAC strategy is dynamically adjusting nodes' duty cycle based on their energy level. For example, nodes with higher residual energy may get a higher duty cycle and transmission priority, while low-energy nodes may stay quiescent longer to recharge \cite{Sangrez.2021,Giannini.2025}. Similarly, ZEDs might use a multi-threshold policy, switching to lower duty cycles when energy falls below certain marks and raising it when energy is abundant  \cite{Sarang.2023}. Some of these aspects are discussed next via an example, wherein the goal is to reduce the AoI metric in a network where the ZEDs must report updates about observed environmental or system events. Note that the AoI is defined as the time elapsed since the generation of the most recently received update packet, and thus is a critical metric for quantifying information freshness in real-time monitoring applications \cite{Kosta.2017,Giannini.2025}.

\begin{example}[Energy-aware MAC for AoI reduction]
Consider a setup with $N$ ZEDs, with energy storage capacity $E_M$, monitoring independent events that occur with probability $p$ every time slot. These are reported to a base station through a common channel. One energy unit arrives, i.e., is harvested, with probability $p'$ each time slot. Upon event detection, the ZED encodes the corresponding information into a data packet and buffers it for transmission. If a packet is already buffered when a new event occurs, the old packet is discarded and replaced by the new one. A transmission with $E$ energy units experiences an erasure probability given by the decreasing function $f(E)$. If several ZEDs transmit in the same time slot and packets are not erased, a collision occurs, causing all transmissions to fail. All packets transmitted and not successfully received are lost, as ZEDs do not attempt retransmissions.  

Two possible protocols and EI acquisition methods are considered: i) ZEDs are fully aware of their energy storage levels at all times (referred to as fully-aware) by using the information sampling approach (cf. Section~\ref{sec:sampling}), wherein EI is acquired at the data packet generation intervals; and ii) ZEDs only know when the energy storage level exceeds a predefined threshold (referred to as partially-aware) by using a single-threshold comparator-based monitoring (cf. Section~\ref{sec:compM}). Furthermore, we consider an energy-blind baseline approach where ZEDs operate without EI. In the case of the fully-aware approach, ZEDs may transmit a buffered packet with probability $f'(E)$, non-decreasing on $E$, using all the stored energy. Meanwhile, in the case of the partially-aware approach, ZEDs wait for the stored energy level to reach the threshold $\delta$ before transmitting, and always consume the threshold-related energy level. In the baseline, the ZED attempts a transmission using a fixed amount of energy $E_t$. If the ZED does not have the required energy, the transmission fails while in progress and the packet is discarded, thus wasting energy in vain.

The fully-aware approach can promote successful transmissions by increasing the transmission chances with energy availability. However, low-energy transmissions are still possible and may be unsuccessful, leading to suboptimal energy usage. Moreover, measuring the storage level before each transmission attempt incurs a potentially high energy cost. In contrast, relying on threshold crossing-based transmission is less energy-intensive, but a proper threshold selection is required. Similarly, the energy-blind approach also requires a proper selection of the fixed energy to be consumed, and it may result in excessive wasted energy if this value is set too high or too many erased packets if it is too low.

\begin{figure}[t!]	\includegraphics[width=\linewidth]{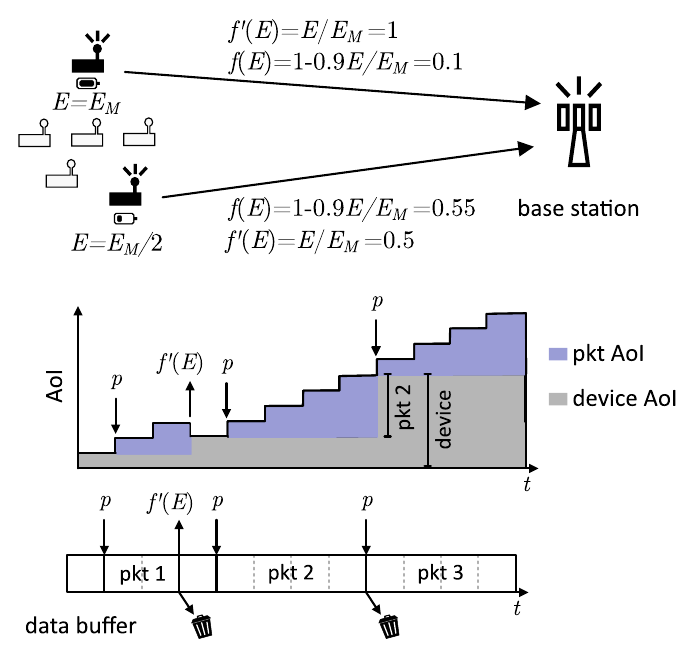}
\caption{System model highlighting the AoI and data buffer processes. Expressions for the erasure probability function $f(E)$ and transmission probability of a buffered packet $f'(E)$ are exemplified.}
\label{fig:energyaware_AoI_system}
\end{figure}
\begin{figure}[t!]
\includegraphics[width=\linewidth]{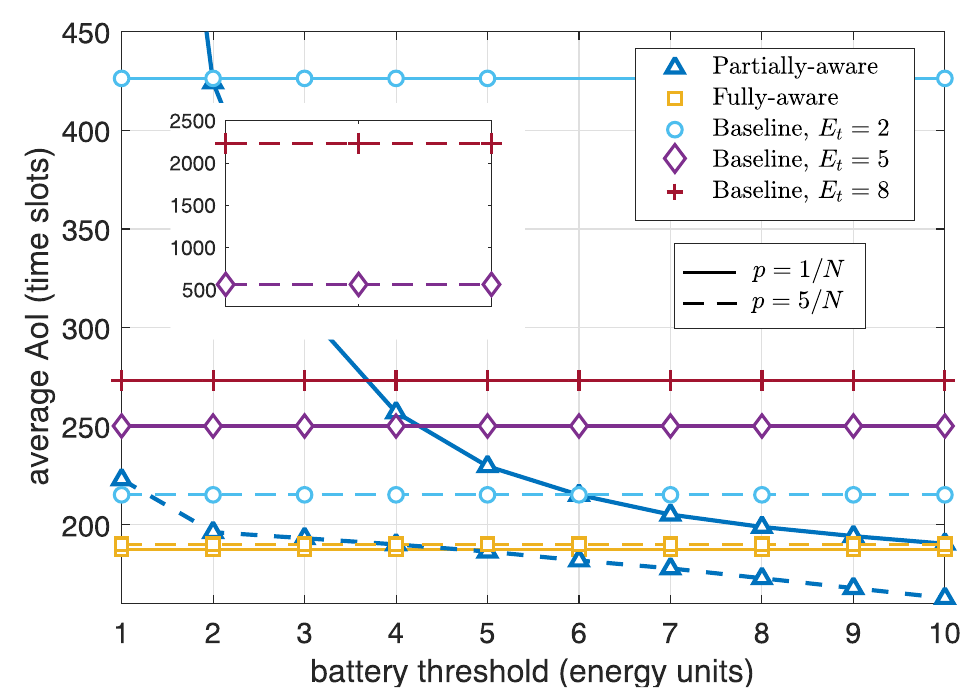}\\%
\includegraphics[width=\linewidth]{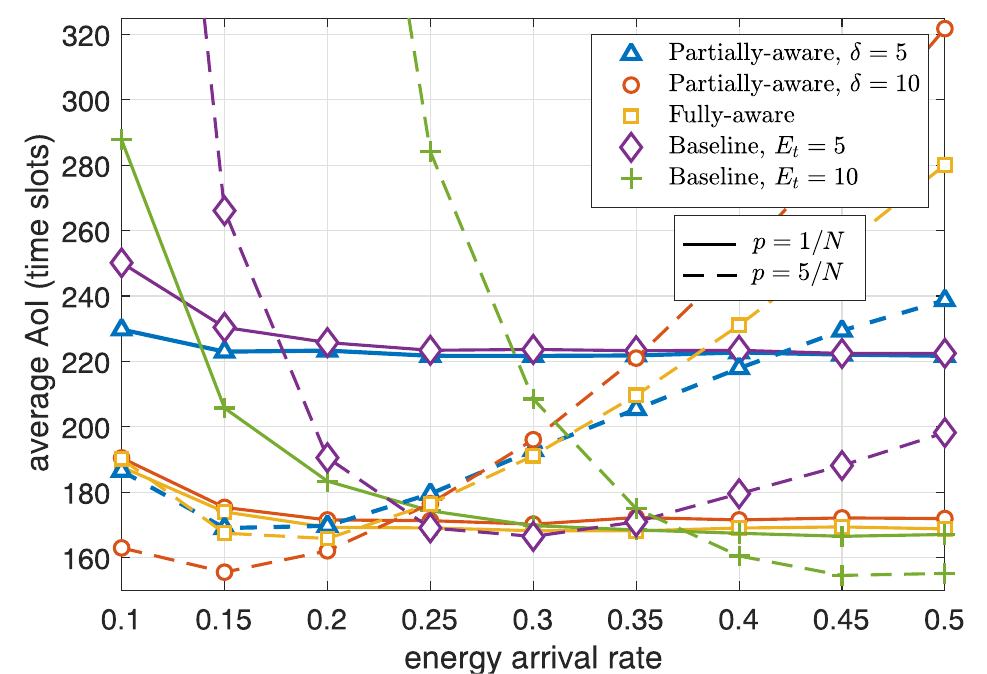}%
\caption{Average AoI as a function of the battery level threshold $\delta$ (top) and energy arrival rate $p'$ (bottom). We consider a scenario with $N=64$ ZEDs, each with an energy buffer of $E_M=10$ units, while setting the probability of events $p=\{1/N, 5/N\}$ and configuring $f(E)$ and $f'(E)$ as shown in Fig.~\ref{fig:energyaware_AoI_system}. For the top figure, the probability of EH in a time slot is set to $p'=0.1$.}
\label{fig:energyaware_AoI}
\end{figure}

Fig.~\ref{fig:energyaware_AoI_system} illustrates the system dynamics, and Fig.~\ref{fig:energyaware_AoI} shows the average AoI as a function of the energy threshold $\delta$ used for the partially-aware approach (top) and the energy arrival rate $p'$  (bottom). Since the energy threshold is not applicable for the fully-aware case and the baseline, their corresponding results appear as straight lines. Observe that the partially-aware setup achieves minimum average AoI when the threshold equals $E_M$ and events are rarer, i.e., lower $p$, resulting in performance similar to that of the fully-aware approach. On the other hand, when events are more frequent, i.e., larger $p$, the partially-aware approach can outperform the fully-aware one if the threshold is properly tuned according to the system parameters. However, modifying the threshold may not be possible in simple ZED implementations, limiting their deployment dynamicity/scalability in practice. Meanwhile, a feedback link from the base station to the ZEDs may be required when threshold optimization is possible. All these aspects require further research.  

 Lastly, in this scenario, the energy-blind baseline performs worse than both energy-aware approaches. As energy arrivals are relatively rare, its oblivious behavior results in many failed transmission attempts, although there are still instances
 where the ZEDs have enough energy for a proper transmission. On the other hand, if energy arrivals are more frequent, having some ZEDs deplete their batteries may be useful, as it may allow other ZEDs to have successful transmissions in the future. 
Ultimately, the choice between approaches depends on whether the ZED can afford to monitor its battery continuously or if it is realistic to provide feedback between users and the base station to fine-tune the threshold. Moreover, depending on the system parameters, such as event occurrence and energy arrival rate, one approach may be preferred instead of the other.
\end{example}

Note that ZEDs with energy-forecasting capabilities, e.g., due to known environmental cycles or using TinyML, can adjust MAC parameters proactively as in \cite{Sarang.2023}.
There is also the option for wake-up radio-based access, where ultra-low-power secondary radios activate the main transceiver only when needed according to a request from a network node, enabling energy-efficient, on-demand communication \cite{Lopez.2023,Guirola.2024,Guirola.2025}. Meanwhile, at the network layer, routing protocols can exploit knowledge of nodes' residual energy, harvesting rate, and link quality to determine reliable paths or even degrade quality-of-service gracefully by lowering data rates or prioritizing critical messages when energy is scarce \cite{Hesam.2024}.

All in all, energy-aware communication protocols across all layers must converge toward designs that avoid but can also tolerate power intermittency, adapt to variable energy contexts, and prioritize robustness over strict determinism.

\subsection{Energy-aware Actuation}
Energy-aware actuation may be as simple as deferring non-critical actuator tasks until energy conditions are favorable \cite{Maeng.2020}. For example, a batteryless e-ink display may postpone screen refreshing, and an irrigation valve or smart lock may queue an actuation event for later, ensuring the action completes without mid-operation power loss. Actuators can be treated as tasks with deadlines and energy costs, and energy availability predictions used to properly set actuation timing \cite{Delgado.2022}. Another approach is modulating the actuation intensity or duration based on real-time or predicted energy availability \cite{Maeng.2020}. For instance, a servo motor may move with reduced torque or sweep angle; while e-paper or LED displays and PWW dimming may reduce brightness when energy is scarce, trading performance for guaranteed completion.

Energy-aware actuation can also target the EH interface itself. Unlike typical PMUs, which optimize energy transfer through electrical regulation and impedance matching, these reconfigurations involve physical or structural adaptations of the transducer or energy-storage reconfigurations. They consume energy but can significantly enhance long-term EH and/or energy consumption efficiency in dynamic environments. For example, vibration harvesters can modulate their mechanical resonance \cite{Mosch.2020,Ibrahim.2021}, solar or thermoelectric harvesters may adapt optical or thermal interfaces to improve energy intake \cite{Maciej.2020}, and RF transducers equipped with multiple antennas and an RF combining circuit can adjust their RX beam direction by actuating phase shifters to align with the most favorable RF energy source, as exemplified next.

\begin{example}[Dynamic RF combining  \cite{Lopez.2022}] 
\begin{figure}[t!]
    \centering
    \includegraphics[width=0.97\linewidth]{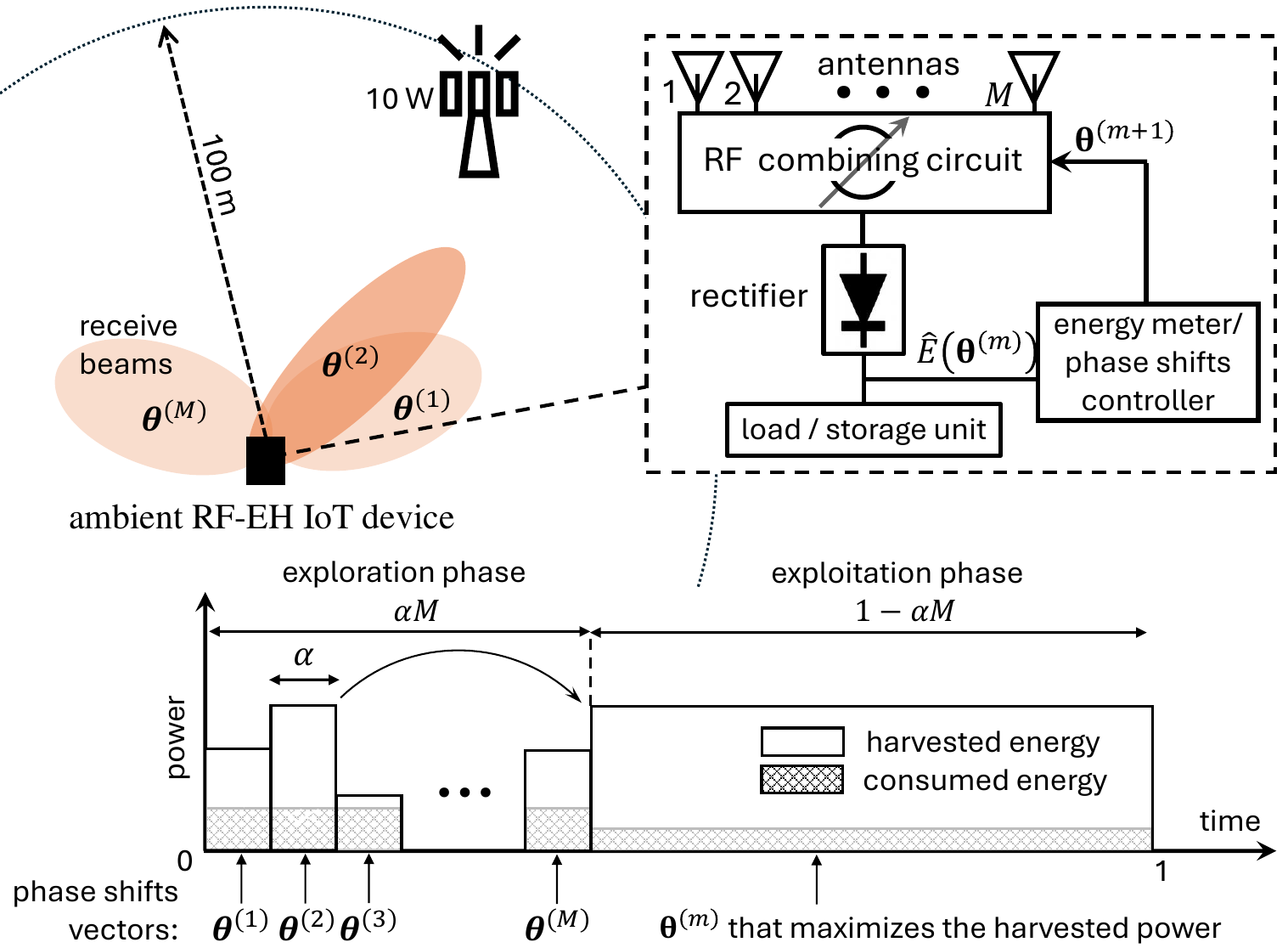}\\
    \includegraphics[width=0.97\linewidth]{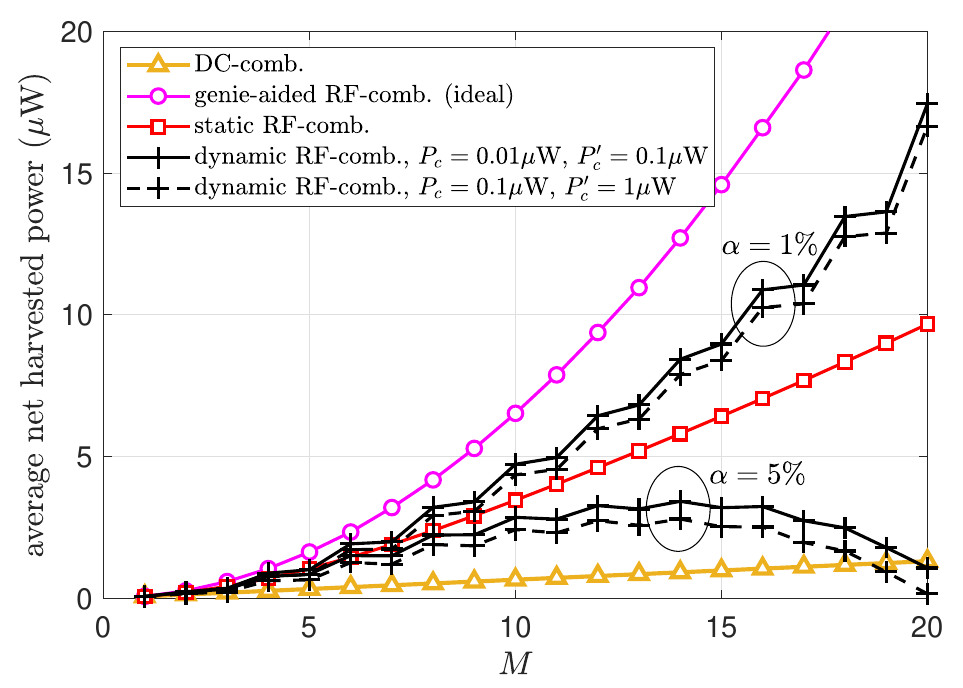}
    \caption{(top) Architecture, configuration protocol, and example setup for dynamic RF combining from ambient RF signals; and (bottom) average net harvested power as a function of the number of antennas for the example setup, including comparisons with DC combining, genie-aided RF combining, and static RF combining benchmarks. 
    In the simulation setup, the RF-EH ZED is illuminated by an ambient RF isotropic source of 10 W, randomly located in a circular area of 100 m radius. Moreover, we assume full line-of-sight channels, a uniform linear array at the RF-EH ZE, $50\%$ EH efficiency, and a log-distance path-loss model with exponent $2.7$ and 40 dB loss at 1 m.}
    \label{fig:dynRF}
\end{figure}

Fig.~\ref{fig:dynRF} illustrates the antenna-rectifier architecture and configuration protocol for dynamic (energy-aware) RF combining from ambient RF signals using $M$ antennas. In an exploration phase, $M$ phase shift configurations $\{\mathbf{\theta^{(m)}}\}$ drawn from a discrete Fourier matrix codebook are tested, while the one providing the greatest DC ETO power is selected for the subsequent exploitation phase. Simulation results in terms of average net harvested power are shown at the bottom of Fig.~\ref{fig:dynRF}. 

As benchmarks, we illustrate the performance of i) DC combining, which uses one rectifier connected to each antenna and thus realizes an omnidirectional RX radiation pattern; ii) genie-aided RF combining, which determines instantaneously and without cost the optimum RF combining configuration, thus providing an upper-bund performance; and static RF combining, which corresponds to a nontunable RF architecture with phase shift configuration $\mathbf{\theta}=[0, \pi,0,\pi,\cdots]$, which provides the widest main high-gain beams \cite{Lopez.2021}. Dynamic RF combining is affected by the power required for tuning each antenna phase shifter ($P_c$) and for EI acquisition after each test phase shift $\mathbf{\theta}^{(m)}$ ($P_c'$), producing EI estimate $\hat{E}\big(\mathbf{\theta}^{(m)}\big)$.

The results reveal the significant influence of the energy consumption of the EI acquisition and actuation circuitry on the potential performance gains of such a dynamic RF combining approach. Indeed, if the exploration phase, wherein several phase shift configurations are tested and their energy output measured, extends too long, RF combining dynamicity is prejudicial and can be outperformed by static RF combining and even DC combining. Meanwhile, if the exploration phase is kept light, both in terms of time and energy consumption, then significant performance gains over feasible benchmarks can be achieved.
\end{example}

Finally, energy storage may be reconfigured to match application-specific energy demands, as proposed in \cite{Colin.2018}. Therein, a combination of SW commands and HW control is used to activate switches that connect or disconnect capacitor banks, ensuring efficient and responsive energy management.
\subsection{Energy-aware Security}\label{sec:sec}
\begin{figure*}[t!]
    \centering
    \includegraphics[width=\linewidth]{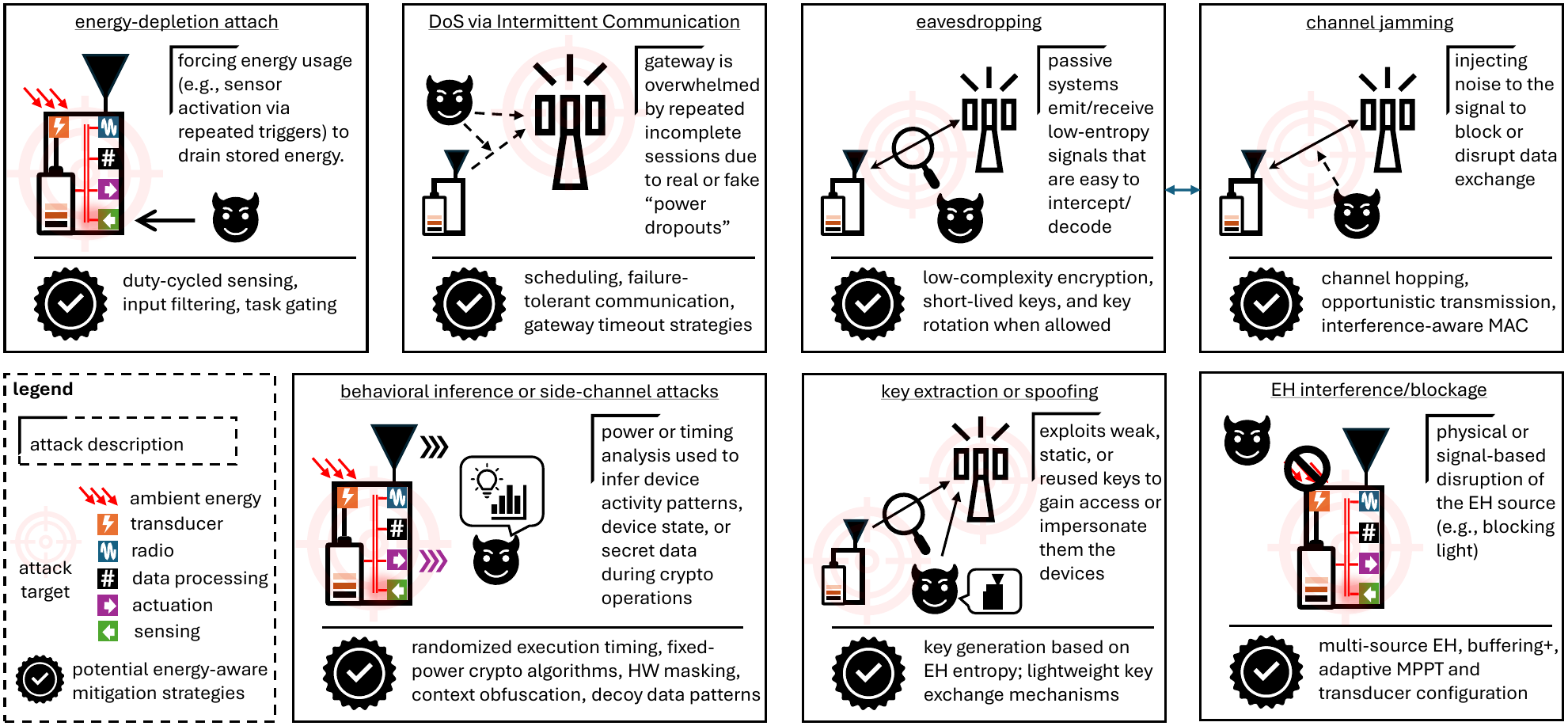}
    \caption{Key attack vectors targeting ZED systems, including affected components, typical mechanisms, and potential energy-aware mitigation strategies.}
    \label{fig:attacks}
\end{figure*}

ZED systems are especially vulnerable to malicious attacks due to their constrained resources, intermittent power supply, and limited ability to execute traditional security mechanisms reliably, calling for tailored security approaches.  Fig.~\ref{fig:attacks} highlights these inherent ZED vulnerabilities\footnote{Refer to \cite{Khan.2022} for detailed discussions on vector attack vectors in IoT networks and related countermeasures. However, note that energy aspects are therein significantly overlooked.} and potential energy-aware mitigation procedures, while Table~\ref{tab:sec} lists typical security tasks as previously discussed in Sections~\ref{sec:comp} and \ref{sec:comm}, their key features, and examples of how energy-awareness can benefit them. Indeed, typical security operations like authentication, encryption/decryption, key exchange, access control, and integrity checking are often crucial \cite{Grossschadl.2007,Khan.2022}. Some of them must execute at system startup (e.g., secure boot), while others are event-driven or embedded within communication processes.

\begin{table*}[t!]
    \centering
     \caption{Typical security tasks and their key ZED related features and potential energy-aware mitigation mechanisms}
    \begin{tabular}{p{3cm}|p{1.5cm}|p{2cm}|p{2.9cm}|p{1.88cm}|p{4.8cm}} \toprule
     \textbf{security task} & \textbf{energy cost} & \textbf{timing flexibility} & \textbf{interrupt tolerance} & \textbf{execution scope} & \textbf{energy-aware mitigation} \\ \midrule
     authentication (e.g., device pairing, challenge-response)  &  medium-high & often tight (real-time) &	low (requires completion to succeed) & per-session or event-driven &	threshold-triggered execution, delay until energy buffer suffices, low-power crypto cores \\ \hdashline
     encryption/decryption (e.g., AES block cipher, ECC encryption) &	medium	& moderate &	moderate (if chunked or buffered) &	per-packet or data block &	algorithm selection (e.g., lightweight ciphers), partial buffering, energy-aware cipher modes \\ \hdashline
     key generation, derivation	& high, low (EH-derived) &	low (often time-sensitive) &	very low (incomplete exchange breaks session) &	session-level &	key generation exploiting EH entropy, key rotation, scheduling \\ \hdashline
     key exchange & high & low & very low & session initiation & opportunistic scheduling, threshold-based handshakes, protocol offloading to edge \\ \hdashline
     integrity MAC checking (e.g., message authentication codes, CRC) & low-medium & high (can be deferred) & high (stateless checks possible) & per-packet or per-frame &  lightweight MAC algorithms (e.g., SipHash), lazy checking, adaptive checking frequency \\ \hdashline
     access/policy control (e.g., conditional resource access, whitelists) &  low–medium &	moderate–high &	high &	event-driven or rule-based &	context-aware policy gating, delayed evaluation, non-volatile policy caching \\ \hdashline
     secure boot, firmware check (e.g., signature verification, hash checks) &	medium–high &	low (startup-only) &	very low (must complete fully)	& at power-up &	optimized signature checks, static hash storage, checkpointed verification stages \\ \bottomrule
    \end{tabular}   
    \label{tab:sec}
\end{table*}

Applicable energy-aware security mechanisms naturally include previously-discussed energy-aware sensing, communication, computation, and actuation tasks to avoid energy-depletion attacks, but also scheduling secure tasks according to energy availability \cite{Ma.2020}. This includes the control logic for secure state management, such as tracking session timeouts, storing partial handshake data, or deciding when to re-authenticate, which must operate reliably across power cycles, often relying on non-volatile memory and energy-aware checkpointing strategies to preserve consistency. In addition, the EH context may be exploited for lightweight key generation. This is promising for enabling symmetric key establishment without interactive protocols, e.g., shared vibration signals harvested via piezoelectric ZEDs may support low-cost, synchronized key generation between co-located ZEDs \cite{Lin.2019}. Whether similar techniques can be extended to other energy sources, like solar, remains an open and promising research direction.
\section{Conclusions: Takeaways \& Challenges Ahead \label{sec:conclusion}}
This work explored the potential and fundamentals behind energy-aware operation for ZEDs. Specifically, we examined the modeling of EI acquisition, task-level behavior, energy availability and usage dynamics, and protocol strategies with a focus on real-world constraints. The main takeaways are:
\begin{itemize}
    \item Theoretical models should reflect actual energy source behavior, storage limitations and corresponding energy evolution dynamics, and load demands to ensure realism. Indeed, even small modeling inaccuracies may result in unworkable protocols as ZEDs are highly constrained and their components' functioning is highly intertwined.
    \item EI may refer to measurements or estimates that characterize a device's energy state(s) or dynamics. This includes, but is not limited to, power input from the energy source, stored energy level (e.g., capacitor voltage or battery SoC), energy consumption rates, predicted future energy availability, and energy trends over time. Decisions about how and when to gather EI must account for trade-offs in cost, accuracy, and protocol impact.
    \item Energy storage characteristics and imperfections interact closely with energy usage architectures and protocol logic, shaping the device’s energy dynamics. 
    \item The energy and timing characteristics of operation tasks dictate how they can be scheduled, adapted, or deferred under intermittent energy conditions. Task characterization may require detailed profiling of energy consumption, execution duration, and responsiveness constraints, as well as understanding how these metrics vary across different operating modes and environmental conditions.
    \item Energy-aware protocols must balance responsiveness with short/long-term energy stability while weighing the overhead of EI acquisition and adapting to energy changes.
\end{itemize}
These are supported by a meticulous literature review, digests from the state-of-the-art, fundamental mathematical frameworks, and in-house examples.

To conclude, Table~\ref{tab:challenges} captures the main challenges pending and research directions, some of which were already identified throughout the paper. These are intended to support future efforts toward more realistic, efficient, and adaptable energy-aware protocol designs for ZED systems.

\begin{table*}[t!]
\caption{Key research gaps and directions in energy-aware ZED protocols}
    \label{tab:challenges}
    \centering
    \begin{tabular}{p{2.2cm}|p{7.8cm}|p{7cm}}\toprule
      \textbf{topic}   & \textbf{open challenges or research gaps} & \textbf{research directions}  \\ \midrule
      EI acquisition modeling &	acquisition overheads and fidelity are often omitted or treated simplistically in theoretical models &	formalize acquisition costs and resolution constraints within energy-aware protocol models; study protocol co-design with EI mechanisms \\ \hdashline
      load and task behavior abstractions & loads and operation tasks are typically treated as opaque or static consumers of energy, without linking their behavior to circuit-level load models  &	investigate how canonical load abstractions (e.g., CR, CI, CP, and hybrids) can model diverse task behaviors and inform power management, scheduling, and protocol scheduling). \\ \hdashline
      battery modeling and integration & battery-equipped ZEDs remain underrepresented in protocol studies and existing models often neglect nonlinear charging dynamics, hysteresis, and long-term degradation effects &	develop lightweight battery models tailored to ZEDs and incorporate them into energy-aware protocol logics \\ \hdashline
      sensitivity of modeling assumptions & lack of systematic analysis on which physical factors related to EH, storage, and consumption dominate system dynamics: models often include/ignore parameters without quantifying their real impact & develop frameworks to quantify the contribution of various factors to modeling accuracy and guidelines prioritizing the inclusion of high-impact parameters according to the scenario \\
      \hdashline
      holistic energy-aware protocols &	energy adaptation is typically confined to a single layer or domain (e.g., MAC, application task scheduling) &	explore multi-layer/domain energy-aware architectures that coordinate adaptation across layers and domains \\ \hdashline
      security in energy-aware systems &	energy-aware ZED protocols largely ignore security implications, such as EI spoofing, denial-of-energy attacks, or manipulation of adaptation logic &	develop lightweight security mechanisms tailored for ZEDs, and explore security–energy trade-offs in adaptive and networked protocols \\ \hdashline
      comparative evaluation of ZED designs & missing systematic studies comparing different harvesters, storage types, or environmental scenarios using consistent metrics and configurations &	conduct realistic, cross-platform evaluations to identify design trade-offs and performance envelopes, helping to mature the conceptual and practical foundations of ZED systems \\ \hdashline
      energy data collection and use &	EI is typically treated as an internal resource, with little attention to its lifecycle, sharing strategies, or security implications when exposed to external systems &	investigate frameworks for secure, efficient energy data sharing (e.g., to networks or digital twins), including data abstraction, local vs. remote use trade-offs, and access control mechanisms \\       \hdashline
      context-aware protocols &  temperature drifts, component aging, and environmental factors are commonly ignored by state-of-the-art protocol logics, while their effects on measurements and energy dynamics may be significant        & extend energy-aware to context-aware protocols by modeling coupled physical effects and internal system states, e.g., thermal-electrical feedback \\   \hdashline
      local–network protocol coupling & energy-aware control is mostly confined to the device side, with limited consideration of how network infrastructure or digital twins can support adaptation &	explore distributed architectures where network-side intelligence complements ZED-local protocols, enabling context-aware coordination, offloading, or predictive adaptation\\
      \bottomrule
    \end{tabular}    
\end{table*}

\bibliographystyle{IEEEtran}
\bibliography{references}

\end{document}